\documentclass[11pt]{article}
\usepackage{graphicx}
\usepackage{amsmath,amsfonts,amssymb}
\usepackage{color}
\usepackage{bbold}
\usepackage{braket}
\usepackage{leftindex}
\usepackage{tikz}
\allowdisplaybreaks
\def\sloppy{\tolerance=100000\hfuzz=\maxdimen\vfuzz=\maxdimen}
\allowdisplaybreaks
\def \beq  {\begin{equation}}
\def \eeq  {\end{equation}}
\def \beqar {\begin{eqnarray}}
\def \eeqar {\end{eqnarray}}
\def\sqr#1#2{{\vcenter{\vbox{\hrule height.#2pt
\hbox{\vrule width.#2pt height#1pt \kern#1pt
\vrule width.#2pt}\hrule height.#2pt}}}}

\mathchardef\mhyphen="2D
\def\la {{\langle}}
\def\ra {{\rangle}}
\def\vx {{\vec x}}
\def\vy {{\vec y}}
\def\vk {{\vec k}}
\def\vf {{\varphi}}

\def\Tr {{\rm Tr}}

\def\tS{{\tilde S}}

\def\tb{{\tilde b}}

\def\rL{{\rm L}}
\def\rR{{\rm R}}

\def\vk {\vec{k}}

\def\vx {{\vec x}}
\def\vz {\vec{z}}
\def\vy{\vec{y}}

\def\vf {{\varphi}}

\def\del {\partial}

\def\a {\alpha}
\def\b {\beta}
\def\e {\epsilon}

\def\C {{\cal C}}

\def\O {{\cal O}}
\def\P {{\cal P}}

\def\half{\textstyle{1\over 2}}

\begin{document}
%%%%%%%%%%%%%%%%%%%%%%%%%%%%%%%%%%%%%%%%%%%%%%%%
%%%%%%%%%%%%%%%%%%%%%%%%%%%%%%%%%%%%%%%%%%%%%%%%
%%%%%%%%%%%%%%%%%%%%%%%%%%%%%%%%%%%%%%%%%%%%%%%%
\fontfamily{bch}\fontsize{11pt}{14.5pt}\selectfont
%\fontfamily{pnb}\fontsize{12pt}{16pt}\selectfont
%\fontfamily{pzc}\fontsize{14pt}{16pt}\selectfont
%\fontfamily{pbk}\fontsize{12pt}{16pt}\selectfont
%\fontfamily{cmr}\fontsize{11pt}{15pt}\selectfont
%\fontfamily{phv}\fontshape{ro}\fontsize{11pt}{14pt}\selectfont
%\fontfamily{ptm}\fontseries{m}\fontshape{r}\fontsize{12pt}{16pt}\selectfont
%\fontfamily{pnc}\fontseries{m}\fontshape{r}\fontsize{11pt}{15pt}\selectfont
%\fontfamily{ppl}\fontseries{m}\fontshape{r}\fontsize{11pt}{15pt}\selectfont
%\usefont{T1}{phv}{m}{it}
%%%%%%%%%%%%%%%%%%%%%%%%%%%%%%%%%%%%%%%%%%%%%%%
\def \CMP {{Commun. Math. Phys.}}
\def \PRL {{Phys. Rev. Lett.}}
\def \PL {{Phys. Lett.}}
\def \NPBProc {{Nucl. Phys. B (Proc. Suppl.)}}
\def \NP {{Nucl. Phys.}}
\def \RMP {{Rev. Mod. Phys.}}
\def \JGP {{J. Geom. Phys.}}
\def \CQG {{Class. Quant. Grav.}}
\def \MPL {{Mod. Phys. Lett.}}
\def \IJMP {{ Int. J. Mod. Phys.}}
\def \JHEP {{JHEP}}
\def \PR {{Phys. Rev.}}
\def \JMP {{J. Math. Phys.}}
\def \GRG{{Gen. Rel. Grav.}}
%%%%%%%%%%%%%%%%%%%%%%%%%%%%%%%%%%%%%%%%%%%%%%%
%%%%%%%%%%%%%%%%%%%%%%%%%%%%%%%%%%%%%%%%%%%%%%%
%%%%%%%%%%%%%%%%%%%%%%%%%%%%%%%%%%%%%%%%%%%%%%%
%%%%%%%%%%%%%%%%%%%%%%%%%%%%%%%%%%%%%%%%%%%%%%%
\begin{titlepage}
\null\vspace{-62pt} \pagestyle{empty}
\begin{center}
%\rightline{CCNY-HEP-18-02}
%\rightline{February 2018}
\vspace{1truein} {\Large\bfseries
Entanglement and Interface Conditions on Fields}\\
\vspace{6pt}
\vskip .1in
{\Large \bfseries  ~}\\
%\vskip .2in
{\Large\bfseries ~}\\
%%%%%%%%%%%%%%%%%%%%%%%%%%%%%%%%%%%%%%%%%%%%%%%%\vspace{.6in}
{\sc Antonina Maj, V.P. Nair}\\
\vskip .2in
{\sl Physics Department,
City College of New York, CUNY\\
New York, NY 10031}\\
\vskip.1in
{\sl The Graduate Center, CUNY\\
New York, NY 10016}\\
\vskip .1in
\begin{tabular}{r l}
{\sl E-mail}:&\!\!\!{\fontfamily{cmtt}\fontsize{11pt}{15pt}\selectfont amaj@gradcenter.cuny.edu}\\
&\!\!\!{\fontfamily{cmtt}\fontsize{11pt}{15pt}\selectfont vpnair@ccny.cuny.edu}\\
\end{tabular}

\vspace{.8in}
%\vspace{1.5in}%\vspace{0.3in}
\centerline{\large\bf Abstract}
\end{center}
We consider the vacuum wave function
of a free scalar field theory in space partitioned into
two regions, with the field obeying Robin conditions (of parameter $\kappa$) on the interface. A direct integration over fields in a subregion is carried out to obtain the reduced density matrix. This leads to a constructive proof of the Reeh-Schlieder theorem. We analyze the entanglement entropy as a function of the Robin parameter $\kappa$. We also consider a specific conditional probability as another measure of entanglement which is more amenable to analysis of the dependence on interface conditions.
Finally, we discuss a direct calculation of correlation functions and how it gives
an alternate route to the reduced density matrix.

\end{titlepage}

%%%%%%%%%%%%%%%%%%%%%%%%%%%%%%%%%%%%%%%%%%%%%%%%
%%%%%%%%%%%%%%%%%%%%%%%%%%%%%%%%%%%%%%%%%%%%%%%%
%%%%%%%%%%%%%%%%%%%%%%%%%%%%%%%%%%%%%%%%%%%%%%%%
%%%%%%%%%%%%%%%%%%%%%%%%%%%%%%%%%%%%%%%%%%%%%%%%
\pagestyle{plain} \setcounter{page}{2}
%%%%%%%%%%%%%%%%%%%%%%%%%%%%%%%%%%%%%%%
%%%%%%%%%%%%%%%%%%%%%%%%%%%%%%%%%%%%%%%
\section{Introduction}
%%%%%%%%%%%%%%%%%%%%%%%%%%%%%%%%%%%%%%%
%%%%%%%%%%%%%%%%%%%%%%%%%%%%%%%%%%%%%%%
Entanglement has long been recognized as the defining characteristic of quantum theory and it continues to be an intriguing theme of much of current research in theoretical physics \cite{reviews}. In the present phase of resurgence of interest, entanglement has been viewed
as perhaps the proper way to characterize topological phases of matter \cite{top-phases}.
The fact that many examples of topological phases have recently been identified has given significant impetus to analyzing the implications as well as different realizations of entanglement.
Quantum computing has been another strong motivation to understand
entanglement in all its ramifications \cite{q-comp}.
Perhaps, the strongest motivation has been the relation between entanglement and gravity. It has long been recognized, at least since the work of Bekenstein and Hawking \cite{bek-hawk}, that entropy does play an important role
in gravity. It is also closely related to the first law of thermodynamics and the Einstein equations for gravity \cite{jacobson}.
The more recent interpretation or identification of this entropy with the entanglement entropy \cite{EE-grav-early, EE-grav-later, EE-grav-string} is suggestive of some intricate relation between quantum mechanics and gravity. 
(Many of the earlier calculations of the entanglement entropy were also motivated by hints of such a relation \cite{Raam}.)
Into this mix, one could add the fact that in
relativistic quantum field theory almost every state shows
entanglement over spacelike separations.
The Reeh-Schlieder theorem
captures the essence of
this property, showing that every state in the full field theory has nonzero overlap with some state constructed in terms of local
observables (defined in some small spacetime region) acting on the vacuum state \cite{RS}.
(This is also related to the fact that
the relevant algebra of local observables is a Type III$_1$ von Neumann algebra \cite{{vN},{vN2},{witten}}.)

Given the intimations of deeper relations among what are {\it a priori} different
concepts, it is clear that the study of entanglement in field theory, even just for the vacuum state, should be very interesting.
And not surprisingly, this has been done with varying degrees of sophistication over the last several years.
Such calculations have focused, for technical reasons, on
path integrals and the replica trick \cite{replica}.
One defines the vacuum wave function $\Psi_0 $ in terms of a path integral
over the time interval $t\in [-\infty, 0]$, the conjugate wave function
$\Psi^*$ can be identified as a similar path integral over the time interval
$[\infty, 0]$. Putting the two together and integrating over fields in
a part of the spatial manifold amounts to the full path integral
over the interval $[-\infty, \infty]$, but with a cut at $t = 0$ corresponding to
the region and field values (which are the arguments of the wave functions) which are not integrated over.
This defines the reduced density matrix for the vacuum state restricted to the region of interest. To facilitate the computation of the R\'enyi entropy, one considers several copies and takes a suitable limit as in the replica trick.
While this is an impressive calculation, we feel that certain physical aspects of the problem are obscured by the calculational procedure.
So the first question we want to consider in this paper is whether
one can get the reduced density matrix by using the
vacuum wave functional on a fixed time-slice and
directly integrating out fields 
in a subregion of space. This should provide a straightforward realization of the Reeh-Schlieder theorem, albeit for the particular field theory under consideration.

A related question is about the influence of boundary conditions
or the conditions obeyed by the fields on the interface or entangling surface.
In most cases of entanglement in field theory, one considers open boundary conditions for the fields on the entangling surface.
However, there are many situations where other choices of interface conditions can be relevant. Consider, for example, the wave function
for the ground state of the electromagnetic field with two parallel conducting plates, the classic geometry
used in Casimir energy calculations.
Clearly, the modes of the fields in $\Psi_0 [\vf ]$, or their normal derivatives,
will vanish on the plates, which is ultimately the reason for the nonzero Casimir energy, and hence any reduced density matrix will be very different from what it is for the vacuum wave function without the plates.
In particular, one would expect, for perfectly conducting plates, no entanglement between physics in the region
between the two plates and the exterior region.
If the plates are replaced by dielectric surfaces, one gets a situation
intermediate between fully open and fully closed interface conditions.
This example shows that it is meaningful and physically relevant
to consider the influence of interface conditions on entanglement and the reduced density matrix.
Ultimately, this question is of the same nature as relating
entanglement and the Einstein equations, since in this latter case, one is considering how entanglement is influenced by geometric data on the interface.
In this paper, we will set up a massless scalar field theory with general Robin 
conditions on the interface and calculate how varying the Robin 
parameter $\kappa$ will affect the reduced density matrix.

In the next section we start with the vacuum wave functional and the construction of modes obeying the appropriate conditions on the interface
(or entangling surface). Integrating out the fields in a subregion and the reduced density matrix are discussed.
We also show how the Reeh-Schlieder theorem is obtained in the particular field theory we have. In section 3 we calculate the entanglement entropy.
The dependence of this quantity on the Robin parameter $\kappa$
is analyzed analytically and numerically for small $\kappa$ and numerically for large $\kappa$. In section 4 we relate the density matrix to correlation functions. In section 5 we consider the conditional probability
of observing a particle anywhere in some region of space given that it was
introduced in some state in the complementary region of space.
We argue that this is also a good measure of entanglement and
analyze its dependence on $\kappa$.
The paper concludes with a short Appendix where we outline how the correlators with the given interface behavior
can be directly calculated. This, along with the discussion in section 4, will give an alternate route to the reduced density matrix.

%%%%%%%%%%%%%%%%%%%%%%%%%%%%%%%%%%%%%%%
%%%%%%%%%%%%%%%%%%%%%%%%%%%%%%%%%%%%%%%
\section{The reduced density matrix}
%%%%%%%%%%%%%%%%%%%%%%%%%%%%%%%%%%%%%%%
%%%%%%%%%%%%%%%%%%%%%%%%%%%%%%%%%%%%%%%
We start with a (1+1)-dimensional field theory on the spatial
interval $[0,L]$, with a partition of this interval into
two segments at $x = b$. The segment $[0,b]$ will be referred to as
the left interval and $[b, L]$ as the right interval.
Additional dimensions transverse to the $x$-axis can be included
without any conceptual or calculational difficulty, so
as a first step we avoid clutter by focusing on the massless (1+1)-dimensional case.
The action for the scalar field is given by
\beq
S = {1\over 2}  \int d^2x \left( {\dot \vf}^2 - (\nabla \vf )^2 \right)
-{1\over 2} \int_{x = b} \!\!dt \, \vf \,\kappa \, \vf
\label{ent1}
\eeq
The second term allows for a proper variational problem with
different boundary/interface conditions. In fact the variation of the action gives
\beq
\delta S = \int \left( - {\ddot \vf} + \nabla^2 \vf \right) \delta \vf
+ \int_{x=b} \left(  \del_n \vf_\rR- \del_n \vf_\rL - \kappa\, \vf  \right) \delta \vf
\label{ent2}
\eeq
where $\del_n \vf_\rL$ and $\del_n \vf_\rR$ denote the normal derivatives
of $\vf$ (i.e., the derivative with respect to $x$ in this case) approaching the interface from the left and right respectively.
The variation (\ref{ent2}) shows that the equations of motion are
\beq
- {\ddot \vf} + \nabla^2 \vf = 0
\label{ent3}
\eeq
with the boundary or interface behavior
\beq
\del_n \vf_\rR - \del_n \vf_\rL = \kappa\, \vf
\label{ent4}
\eeq
For $\kappa \rightarrow 0$, we get the Neumann or open interface conditions at $x = b$ while for $\kappa \rightarrow \infty$, we get the Dirichlet behavior
$\vf = 0$ at $x =b$.
In the following we will take $\kappa$ to be positive. In the case of
negative values for $\kappa$, there can be bound states 
which are localized near the interface. 
Although we do not analyze this case, we will comment on this
possibility in the last section of the paper. 
%%%%%%%%%%%%%%%%%%%%%%%%%%%%%%%%%%%%%%%
%%%%%%%%%%%%%%%%%%%%%%%%%%%%%%%%%%%%%%%
\subsection{Modes on a partitioned interval}
%%%%%%%%%%%%%%%%%%%%%%%%%%%%%%%%%%%%%%%
%%%%%%%%%%%%%%%%%%%%%%%%%%%%%%%%%%%%%%%
The vacuum wave function will be given in terms of the eigenvalues of
$-\nabla^2$. Therefore, we consider the eigenvalue equation
\beq
- \nabla^2 \, u = \omega^2 \, u
\label{ent5}
\eeq
The solution is given by
\beq
u = \begin{cases}
A \, \sin \omega x\hskip .5in& x\in [0,b]\\
B\, \sin \omega (L - x)& x\in [b, L]\\
\end{cases}
\label{ent6}
\eeq
for constants $A$, $B$. We have chosen the boundary condition
$u = 0$ at $x = 0$ and at $x = L$ for simplicity.
We take $L \rightarrow \infty$ at the end, so the results will not be
sensitive to the conditions at $x = 0, L$.
We need continuity of the field at $x = b$; this relates the constants $A$ and
$B$ and we get
\beq
u =  F \begin{cases}
 \sin \omega (L-b)\, \sin \omega x\hskip .5in& x\in [0,b]\\
 \sin \omega b\, \sin \omega (L - x)& x\in [b, L]\\
\end{cases}
\label{ent7}
\eeq
with a normalization constant $F$. The interface condition (\ref{ent4})
gives
\beq
\omega\, \sin \omega L + \kappa\, (\sin \omega b) \sin \omega(L-b) = 0
\label{ent8}
\eeq
This determines the eigenvalues $\omega^2$ in (\ref{ent5}).
Notice that for $\kappa = 0$ (open interface), we get $\sin \omega L = 0$,
as expected for modes on the full interval $[0,L]$. For $\kappa \rightarrow
\infty$, (\ref{ent8}) gives $(\sin \omega b ) \sin \omega(L-b) = 0$ allowing for
modes with $\omega = n\pi/b$ on the left and with $\omega = m \pi/(L-b)$
on the right, $m$, $n$ being integers. The normalization condition
$\int dx\,u^* u =1$ determines $F$ as
\beq
F = \sqrt{2} \left[ b \sin^2 \omega (L-b) + (L-b) \sin^2\omega b \,+{1\over \kappa}\sin^2 \omega L  \right]^{-{\half}}
\label{ent9}
\eeq
For $\kappa = 0$ it is easy to see that, combining
(\ref{ent7}), (\ref{ent9}),
\beq
u_N (x) = \sqrt{2\over L} (- \cos \omega_N L ) \, \sin \omega_N x
= \sqrt{2\over L} (- 1)^{N-1} \, \sin \omega_N x,
\hskip .3in \omega_N = {N \pi \over L}
\label{ent10}
\eeq
For $\kappa \rightarrow \infty$ we find two sets of modes
\beq
u_n (x) = \begin{cases}
\sqrt{2\over b} \, \sin \omega_n x \hskip .5in & x\in [0,b]\\
0 & x\in [b,L]\\
\end{cases}
\label{ent11}
\eeq
with $\omega_n = n \pi /b$ and
\beq
u_m (x) = \begin{cases}
0 & x\in [b,L]\\
\sqrt{2\over L-b} \, \sin \omega_m(L- x) \hskip .5in & x\in [b,L]\\
\end{cases}
\label{ent12}
\eeq
with $\omega_m = m \pi/(L-b)$. The modes defined by (\ref{ent7}),
(\ref{ent8}) for arbitrary nonzero $\kappa$ smoothly interpolate
between the two sets (\ref{ent10}) and (\ref{ent11}), (\ref{ent12}).
The set of values for $\omega$ for finite and nonzero $\kappa$ will split
into the two sets $\{ n\pi /b\}$ and $\{ m \pi /(L-b) \}$
as $\kappa \rightarrow \infty$.
By direct integration, it is also easy to check that
\beq
\int dx \, u^*_\omega (x) \, u_{\omega'} (x) = 0
\label{ent13}
\eeq
for $\omega \neq \omega'$. The modes defined by 
(\ref{ent7}), (\ref{ent8}) form an orthonormal set. In terms of these modes, the field
$\vf$ has the expansion
\beq
\vf (x) = \sum_N c_N \, u_N (x), \hskip .3in
c_N = \int dx\, u^*_N (x) \vf (x)
\label{ent14}
\eeq
Using this expansion and (\ref{ent4}), the action can be written as
\beq
S = {1\over 2} \sum_N \left( {\dot c}_N {\dot c}_N - \omega_N^2 c_N^2
\right)
\label{ent15}
\eeq
The Hamiltonian takes the form
\beq
H = {1\over 2} \sum_N \left( {\dot c}_N {\dot c}_N + \omega_N^2 c_N^2
\right) = {1\over 2} \sum_N \left( - {\del^2 \over \del c_N^2} + \omega_N^2 c_N^2 \right)
\label{ent16}
\eeq
The ground state wave function or the vacuum wave function is thus
given by
\beqar
\Psi_0 [\vf] &=& {\cal N} \, \exp\left( -{1\over 2} \sum \omega_N c_N^2\right)
\nonumber\\
&=& {\cal N} \, \exp\left( -{1\over 2}  \int \vf (x) \, \Omega (x, x')  \, \vf (x')\right)
\label{ent17}\\
\Omega (x, x') &=& \sum \omega_N u^*_N (x) u_N (x')
\nonumber
\eeqar
where ${\cal N}$ is a normalization factor.
The metric on the space of fields is given by
$ds^2 = \sum_N dc_N dc_N$, so that the measure of integration for the
normalization of the wave function is $d\mu = \prod_N dc_N$.
%%%%%%%%%%%%%%%%%%%%%%%%%%%%%%%%%%%%%%%
%%%%%%%%%%%%%%%%%%%%%%%%%%%%%%%%%%%%%%%
\subsection{Integrating over modes in a subregion}
%%%%%%%%%%%%%%%%%%%%%%%%%%%%%%%%%%%%%%%
%%%%%%%%%%%%%%%%%%%%%%%%%%%%%%%%%%%%%%%
We now consider restricting the fields to left and right intervals.
Towards this, we define the modes
\beq
v_N (x) = \begin{cases} 
u_N (x) \hskip .2in & x\in[0,b]\\
0 &x\in [b,L]\\
\end{cases}
\label{ent18}
\eeq
\beq
w_N (x) = \begin{cases}
0&x\in[0,b]\\ 
u_N (x) \hskip .2in & x\in[b,L]\\
\end{cases}
\label{ent19}
\eeq
These modes, as defined, are not normalized. We also define an orthonormal set for each interval, which are essentially the $\kappa \rightarrow \infty$ limits as given in (\ref{ent11}), (\ref{ent12}). We will use $\{ v_a\}$ to denote modes
defined on $[0,b]$ and $\{w_\a\}$ to denote modes on $[b,L]$.
Thus
\beqar
v_a (x) &=& \sqrt{2\over b}\,\sin \left({a \pi x \over b}\right)\nonumber\\
w_\a (x) &=& \sqrt{2\over L-b}\, \sin \left({\a \pi (L-x) \over(L-b)}\right)
\label{ent20}
\eeqar
The modes in (\ref{ent18}), (\ref{ent19}) can be expanded
in terms of these as
\beqar
v_N (x) &=& S_{N a} \, v_a (x), \hskip .3in
S_{N a} = \int_0^b v_a \, u_N\nonumber\\
w_N (x) &=& \tS_{N \a} w_\a(x), \hskip .3in
\tS_{N \a} = \int_b^L w_\a \, u_N
\label{ent21}
\eeqar
By virtue of the orthonormality and completeness for the modes
$u_N$ over the full interval $[0,L]$, we get
\beqar
&&\sum_N S_{N a} S_{N b} = \delta_{ab}, 
\hskip .2in
\sum_N \tS_{N \a} \tS_{N \b} = \delta_{\a \b}, \hskip .2in
\sum_N S_{N a} \tS_{N \a} = 0\nonumber\\
&&\sum_a S_{N a} S_{M a} + \sum_\a \tS_{N \a} \tS_{M \a} = \delta_{NM}
\label{ent22}
\eeqar
From the definitions (\ref{ent18}), (\ref{ent19}), we also have, for the
full interval $[0,L]$,
\beq
u_N (x) = S_{N a} v_a (x) + \tS_{N \a} w_\a(x)
\label{ent23}
\eeq
Using (\ref{ent14}), we can now write
\beqar
c_N &=& \int_0^L  u_N (x)\, \vf (x) = S_{N a} b_a + \tS_{N \a} \tb_\a\nonumber\\
b_a &=& \int_0^b  v_a(x) \, \vf (x), \hskip .2in
\tb_\a = \int_b^L  w_\a(x) \vf (x)
\label{ent24}
\eeqar
It is also easy to check that, by virtue of (\ref{ent22}), 
\beq
ds^2 = dc_N dc_N = db_a db_a + d\tb_\a d \tb_\a
\label{ent25}
\eeq
so that the integration measure splits up as
$d\mu = \prod db_a \prod d\tb_\a$.

Writing $c_N$ as in (\ref{ent24}), we get
\beqar
c_N \,\omega_N \, c_N &=& b_a (\Omega_{\rm L})_{ab} b_b
+ \tb_\a (\Omega_{\rm R})_{\a \b} \tb_\b
+ b_a (\Omega_{\rm LR})_{a \a} \tb_\a
+ \tb_\a (\Omega_{\rm LR})^{\rm T}_{\a a} b_a
\label{ent26}\\
(\Omega_{\rm L} )_{ab}&=& S_{Na} S_{N b} \omega_N\nonumber\\
(\Omega_{\rm R} )_{\a \b}&=& \tS_{N\a} \tS_{N \b} \omega_N
\label{ent27}\\
(\Omega_{\rm LR})_{a \a} &=& S_{Na} \tS_{N\a} \omega_N
\nonumber
\eeqar
Notice that $\Omega_{\rm L}$ and $\Omega_{\rm R}$ are real symmetric matrices. Integration over the $\tb$'s will correspond to integrating out the fields in the interval $[b,L]$. Thus the reduced density matrix for the left subregion is given by
\beqar
\rho_{\rm red} (b, b') &=& \bra{b} \rho_{\rm red} \ket{b'} =
\int [d\tb] \, \Psi_0^* (b, \tb ) \Psi_0 (b', \tb)\nonumber\\
&=& {\cal N}^2 \int [d\tb]\, e^{-{\half} b\, \Omega_{\rm L} b - {\half} b' \Omega_{\rm L} b' } e^{ - \tb\, \Omega_{\rm R} \tb - b\, \Omega_{\rm LR} \tb -
\tb\, \Omega^{\rm T}_{\rm LR} b'}
\label{ent28}
\eeqar
This reduced density matrix is defined in the basis where $b_a$ are diagonal.
Carrying out the integration over the $\tb$'s,
\beq
\rho_{\rm red} (b, b') = {\cal N}' \, e^{-{\half} b \left( \Omega_{\rm L} - \sigma \right) b - {\half} b' \left(\Omega_{\rm L} - \sigma \right) b' + b \sigma b' }
\label{ent29}
\eeq
where
\beqar
\sigma_{ab} &=& {\half} (\Omega_{\rm LR})_{a\a}\, (\Omega_{\rm R})_{\a \b}^{-1} \, (\Omega_{\rm LR})_{\b b}^{\rm T} \nonumber \\
{\cal N}' &=& \left( \det{(\Omega_{\rm L} - 2 \sigma) / \pi} \right)^{\half}
\label{ent30}
\eeqar
It is useful to change the basis $\{b_a\}$, so that the reduced density matrix depends only on one matrix. Transforming the basis by $b_a \rightarrow \hat{b}_a=(\Omega_{\rm L})^{1/2}_{ab} b_{b}$,
the reduced density matrix will be given by
\beqar
\rho_{\rm red} (\hat{b}, \hat{b'}) &=& \left( \det{(1 - 2 \Sigma) / \pi} \right)^{\half} \, e^{-{\half} \hat{b} \left(1 - \Sigma \right) \hat{b} - {\half} \hat{b'} \left(1 - \Sigma \right) \hat{b'} + \hat{b} \Sigma \hat{b'} }, \nonumber \\
\Sigma_{ab} &=& {\half} (\Omega_{\rm L})_{ac}^{-\half} \, (\Omega_{\rm LR})_{c\a}\, (\Omega_{\rm R})_{\a \b}^{-1} \, (\Omega_{\rm LR})_{\b d}^{\rm T} \, (\Omega_{\rm L})_{db}^{-\half}
\label{ent31}
\eeqar
The eigenvalues of the reduced density matrix, and hence the entanglement entropy, will be functions of the eigenvalues of $\Sigma$. 
%%%%%%%%%%%%%%%%%%%%%%%%%%%%%%%%%%%%%%%
%%%%%%%%%%%%%%%%%%%%%%%%%%%%%%%%%%%%%%%
\subsection{Diagonalizing the reduced density matrix}
%%%%%%%%%%%%%%%%%%%%%%%%%%%%%%%%%%%%%%%
%%%%%%%%%%%%%%%%%%%%%%%%%%%%%%%%%%%%%%%
To find how exactly the eigenvalues of the reduced density matrix depend on $\Sigma$ it is worth starting from a diagonal basis for $\rho_{\rm red}$ and working backwards. 
For this purpose it is useful to introduce a particle basis $\{ \ket{ n_1 n_2 \cdots} \}$. A diagonal matrix with unit trace will have the form
\beq
\rho_{\rm red} = \det{(1-\Lambda)} \sum_{n_1 = 0}^{\infty} \cdots \left( \prod_i \, \Lambda_ i^{n_i} \, \ket{ n_1 n_2 \cdots}\bra{n_1 n_2 \cdots} \right)
\label{ent32}
\eeq
where $\Lambda$ is a diagonal matrix with eigenvalues $\{\Lambda_i \}$. One can read off the eigenvalues of this matrix as $ \det{(1-\Lambda)} \prod_i \Lambda_i^{n_i}$.

The particle states can be written in terms of creation operators in the standard way: $\ket{n_1 n_2 \cdots} = \prod_i \, {1 \over \sqrt{ n_i !}} (B_i^{\dagger})^{n_i} \ket{0}$, where $[ B_i, B_j^{\dagger} ] = \delta_{ij}$.
In a functional basis $\{\ket{\phi}\}$ the matrix in (\ref{ent32}) will have the form
\beq
\rho_{\rm red} (\phi, \phi') = \bra{\phi} \rho_{\rm red} \ket{\phi'} = \det{(1-\Lambda)} \, \exp \left(B^{\dagger}[\phi] \, \Lambda \, B^{\dagger}[\phi']\right) \, \Psi_0 [\phi] \Psi^*_0 [\phi']
\label{ent33}
\eeq
where we used the fact that the wave functions for the
states are real for a real scalar theory, i.e.,  $\braket{\phi| n_1 n_2 \cdots} = \braket{n_1 n_2 \cdots | \phi}$. The annihilation and creation operators and the vacuum state in this basis are given by
\beqar
B_i [\phi] &=& {1 \over \sqrt{2}} \left[ (\omega)^{\half}_{ij} \phi_j + (\omega)^{-\half}_{ij} {\del \over \del \phi_j} \right] \nonumber\\
B^{\dagger}_i [\phi] &=& {1 \over \sqrt{2}} \left[ (\omega)^{\half}_{ij} \phi_j - (\omega)^{-\half}_{ij} {\del \over \del \phi_j} \right] 
\label{ent34}\\
\Psi_0 [\phi] &=& (\det{\omega / \pi})^{1\over 4} \, e^{-{\half} \phi \omega \phi} \nonumber
\eeqar
where $\omega$ is some real symmetric matrix. Choosing $\omega = \left({ 1 - \Lambda \over 1 + \Lambda}\right)$, and using the fact that
\beq
e^{V \cdot B^{\dagger}[\phi]} e^{-{\half} \phi \omega \phi} = e^{-{\half} \phi\omega \phi \, +\, \sqrt{2}\, V \cdot \sqrt{\omega}\phi \,- {\half} V^2}
\label{ent35}
\eeq
one finds that the matrix in (\ref{ent33}) has the form
\beqar
\rho_{\rm red} (\phi, \phi') &=& \left( \det{(1 - 2 \Sigma) / \pi} \right)^{\half} \, e^{-{\half} \phi \left(1 - \Sigma \right) \phi - {\half} \phi' \left(1 - \Sigma \right) b\phi' + \phi \Sigma \phi' } \nonumber \\
\Sigma &=& {2\Lambda \over (1+\Lambda)^2 }
\label{ent36}
\eeqar
which is the same form as our reduced density matrix in (\ref{ent31}). We can identify the basis $\{ \ket{\phi}\}$ as our basis for the left modes $\{\ket{\hat{b}} \}$, and the matrix $\Sigma$ as the one defined in (\ref{ent31}). 
Solving for $\Lambda$ in terms of $\Sigma$,\footnote{There are two possible solutions: $\Lambda = 2\Sigma / (1\pm \sqrt{1-2\Sigma})^2$. 
However, only the positive solution ensures that $\omega$ is positive and, hence, that $\Psi_0$ is normalizable.} we conclude that the eigenvalues of the reduced density matrix in (\ref{ent31}) are
\beq
\Big\{ \det{(1-\Lambda)} \, \prod_{i} \, \Lambda_i^{n_i} \, \Big| \, n_i = 0, 1, 2 \cdots \Big\}, \hskip .3in
\Lambda = { 2 \Sigma \over (1 + \sqrt{1- 2\Sigma} )^2}
\label{ent37}
\eeq
We started this section with the assumption that $\Lambda$ is diagonal, whereas $\Sigma$ in general is a non-diagonal matrix. However, we can easily rotate the particle basis to get $\rho_{\rm red}$ in terms of a non-diagonal $\Lambda$. 
The important point is that, given (\ref{ent37}), we have an explicit relation between the eigenvalues of $\rho_{\rm red}$ and the eigenvalues of the matrix $\Sigma$. 

Looking back at (\ref{ent34}) we can view the creation operators $B^{\dagger}$ as generating the excited states for the theory defined on the left interval $[0,b]$. 
In terms of our original left modes $\{ b_a \}$ defined in (\ref{ent24}), the states diagonalizing $\rho_{\rm red}$ (modulo some orthogonal transformations) are given by
\beqar
\Psi_0[b] = {\cal N} e^{-{\half} b \tilde{\omega}_{\rm L} b}, \hskip .3in \Psi_{n}[b] = \prod_a {1 \over \sqrt{n_a!}} (B_a^{\dagger})^{n_a} \Psi_0 [b] \nonumber\\
B^{\dagger}_{a}[b] = {1 \over \sqrt{2}} \left[ (\tilde{\Omega}_{\rm L})^{\half}_{ab} b_b - (\tilde{\Omega}_{\rm L})^{-\half}_{ab} {\del \over \del b_b} \right]
\label{ent38}
\eeqar
where $\tilde{\omega_{\rm L}} = \sqrt{ \Omega^2_{\rm L} - 2 \Omega_{\rm L} \Sigma \Omega_{\rm L}}$. Moreover, from (\ref{ent32}) it is clear that all excited states will have non-zero overlap with $\rho_{\rm red}$ as long as all the eigenvalues of $\Lambda$, and hence $\Sigma$, are non-zero.
In fact, in the next section we will show that this is indeed the case for $\kappa = 0$, i.e., for the open boundary conditions at the interface. It will thus provide an explicit realization of the Reeh-Schlieder theorem for the scalar field.

%%%%%%%%%%%%%%%%%%%%%%%%%%%%%%%%%%%%%%%
%%%%%%%%%%%%%%%%%%%%%%%%%%%%%%%%%%%%%%%
\subsection{Proving the Reeh-Schlieder theorem}
%%%%%%%%%%%%%%%%%%%%%%%%%%%%%%%%%%%%%%%
%%%%%%%%%%%%%%%%%%%%%%%%%%%%%%%%%%%%%%%
The set of all states of the form $A_1 A_2 \cdots \ket{0}$ where $A_i$ are elements of the algebra of local operators is dense in the full Hilbert space of the field theory. This is the Reeh-Schlieder theorem.
We can now see how this arises in our framework by considering operators defined on the left interval $[0,b]$. Since the correlators of such local operators are defined by
\beq
\bra{0} A_1 A_2 \cdots \ket{0} = \int \Psi_0^* \, A_1 A_2 \cdots \Psi_0
= \Tr (\rho_{\rm red} A_1 A_2 \cdots),
\label{ent39}
\eeq
the theorem essentially states that $\rho_{\rm red}$ has no zero modes.
From (\ref{ent37}), we see that this is equivalent to the statement that $\Sigma_{ab}$ has no zero modes.
We will now show that this is indeed the case. Suppose $\phi_a$ is a zero mode for
$\Sigma_{ab}$. From the form of $\Sigma_{ab}$ given in
(\ref{ent31}), notice that, in matrix notation,
\beq
\Sigma = A^T A, \hskip .3in
A^T = {1 \over \sqrt{2}}( \Omega_{\rm L} )^{-\half}\, \Omega_{\rm LR} \, (\Omega_{\rm R})^{-\half}
\label{ent40}
\eeq
Thus, $\phi_a \Sigma_{ab} \phi_b = (A\phi)_a (A \phi)_a$, showing that
a zero mode for $\Sigma$ should satisfy $A \phi = 0$.
In turn, since $(\Omega_{\rm L})^{-\half}$ and $(\Omega_{\rm R})^{-\half}$ are invertible matrices, this means that a zero mode $\phi_a$ should obey
\beq
\sum_{a}(\Omega_{\rm LR})^{\rm T}_{\a a} \phi_a = \sum_{N,a} \tS_{N\a} \omega_N \,S_{N a} \phi_a
= 0
\label{ent41}
\eeq
Our aim is to show that the only solution to this equation for the free interface case is $\phi_a = 0$. The free or open interface
corresponds to $\kappa = 0$; it is the Poincar\'e invariant
case for which the theorem was originally developed.

First of all, given that $\sum_N \tilde{S}_{\alpha N} S_{Na} = 0$ from (\ref{ent22}), it follows that
\beq
\sum_{N,a} \tilde{S}_{N \a} S_{N a} \phi_a = 0
\label{ent42}
\eeq
Multiplying (\ref{ent41}) and (\ref{ent42}) by the right modes $w_\a(x)$ in (\ref{ent20}), we get
\beqar
\sum_{N,a} u_N(x) \, \omega_N S_{Na} \phi_a = 0 \nonumber\\
\sum_{N,a} u_N(x) S_{Na} \phi_a = 0
\label{ent43}
\eeqar
in the right subregion $x \in [b, L]$, where the modes
$v_a (x)$ have no support. We can rewrite these equations as
\begin{align}
\sum_N \xi_N \,\omega_N\, u_N(x) &= 0, \hskip .3in x\in [b,L]
\label{ent44}\\
\sum_N \xi_N  u_N(x) &= 0, \hskip .3 in x \in [b,L]
\label{ent45}
\end{align}
where $\xi_N = \sum_a S_{Na} \phi_a$.

For $\kappa = 0$ we can take the modes to be $u_N(x) = \sqrt{2/ L} \sin \left( N \pi x  / L \right).$\footnote{This is the same as in (\ref{ent10}) up to an irrelevant factor of $(-1)^{N-1}$. This factor could also be included in the definition of $\xi_N$.}
Since (\ref{ent45}) holds for all $x$ in the right subregion, we can differentiate it with respect to $x$ and obtain the relation
\beq
\sum_N \xi_N \, \omega_N \, \cos \left( N \pi x / L \right) = 0, \hskip .3in x\in [b,L]
\label{ent46}
\eeq
Combining this with (\ref{ent44}), we see that $\xi_N$ must obey
\beq
\sum_N \xi_N \,\omega_N \, e^{i N \pi x /L} = 0, \hskip.3in x\in [b,L]
\label{ent47}
\eeq
It is now easy to see that there is no solution to this equation for
$\xi_N$ except $\xi_N = 0$ (given that $\omega_N = N \pi /L$ are all non-zero) by the standard arguments 
of analytic continuation. For the sake of completeness, we will briefly outline this argument.
Writing $z = e^{i \pi x /L}$, consider the function
\beq
F(z) = \sum_N \xi_N \,\omega_N \,z^N
\label{ent48}
\eeq
$F(z)$ is analytical for $\vert z \vert < 1 $ and for a small range of $\vert z \vert >1$. This can be seen as follows. 
We are interested in zero modes $\phi$ of finite action.
Disregarding right modes in (\ref{ent24}), the action for $\phi$ is finite if
\beq
\sum_N \omega_N^2 \left(S_{Na} \phi_a\right)\left(S_{Nb} \phi_b\right) < \infty
\label{ent49}
\eeq
Given the definition of $\xi_N$, this means that $\sum_N \xi_N^2 \omega_N^2 < \infty$.
The convergence of this series implies that
\beq
\lim_{N \rightarrow \infty} \left\vert {\xi_{N+1} \,\omega_{N+1} \over \xi_N \, \omega_N }\right\vert < {1 \over r}, \hskip .3in r>1
\label{ent50}
\eeq
where $r$ can be infinitesimally close to $1$, i.e., $r = 1 +\e$.

On the other hand, from (\ref{ent48}), $F(z)$ converges absolutely in the region defined by
\beq
\lim_{N \rightarrow \infty} \left\vert {\xi_{N+1} \,\omega_{N+1} \over \xi_N \,\omega_N }\right\vert < {1 \over \vert z \vert}
\label{ent51}
\eeq
Equation (\ref{ent50}) then guarantees convergence for $\vert z \vert < r$. Thus, $F(z)$ is
an analytic function in the region given by $\vert z \vert < 1+ \e$. 
Equation (\ref{ent47}) then says that this function should vanish for the line segment $x \in [b, L]$, 
or $\theta \in [\pi b/L, \pi]$ for $z = e^{i \theta}$.
Hence, $F(z)$ has to vanish everywhere in the domain $\vert z \vert < 1 +\e$.
In particular, we have $\sum_N \xi_N \, \omega_N \sin \left(N \pi x / L \right) =0$ for all $x \in [0, L]$.
Since $\omega_N$ are non-zero, this shows that all $\xi_N = 0$, since $\{ \sin\left(N \pi x / L \right) \}$ form a complete set. 
Hence, $\phi_a$ must be zero.
We conclude that
there are no zero modes for
$\Sigma_{ab}$. This demonstrates the realization of the Reeh-Schlieder theorem
for our case of a scalar field.

A couple of comments are in order at this point.
First of all, our argument is entirely based on the wave function
and equal-time physics. Analytic continuation in the time-variable, which depends on the Hamiltonian having a lower bound,
is not used. Nevertheless, the nature of the Hamiltonian 
does play a role via the vacuum wave function.
The analytic continuation we have used depends on having the exponentials
in (\ref{ent47}), which required both (\ref{ent44}) and (\ref{ent46}).
The second of these equations, obtained from (\ref{ent45}), depended
crucially on having just $\omega_N$ in 
(\ref{ent44}). This is traceable to $\omega_N$ being the kernel in the exponent of
the vacuum wave function in (\ref{ent17}), a form determined by the
Hamiltonian.
The nonlocality of the kernel $\Omega (x, x')$ in $\Psi_0[\vf]$ is thus 
crucial for our argument.
If we had a constant, or a local operator such as $\omega^2_N$, in place of $\omega_N$ in
(\ref{ent17}), equation (\ref{ent47}) would not follow and
the analytic continuation argument would not be obtained.

%%%%%%%%%%%%%%%%%%%%%%%%%%%%%%%%%%%%%%%
%%%%%%%%%%%%%%%%%%%%%%%%%%%%%%%%%%%%%%%
\section{Entanglement and interface conditions}
%%%%%%%%%%%%%%%%%%%%%%%%%%%%%%%%%%%%%%%
%%%%%%%%%%%%%%%%%%%%%%%%%%%%%%%%%%%%%%%
The degree of entanglement will depend on the interface conditions entirely through the dependence of the matrix $\Sigma$ on $\kappa$, where $\Sigma$ is defined in (\ref{ent31}). 
Before going into explicit calculations of $\Sigma$, it is worth working out how the entanglement entropy depends on $\Sigma$.
%%%%%%%%%%%%%%%%%%%%%%%%%%%%%%%%%%%%%%%
%%%%%%%%%%%%%%%%%%%%%%%%%%%%%%%%%%%%%%%
\subsection{Entanglement entropy}\label{entropy}
%%%%%%%%%%%%%%%%%%%%%%%%%%%%%%%%%%%%%%%
%%%%%%%%%%%%%%%%%%%%%%%%%%%%%%%%%%%%%%%
Since the eigenvalues of the reduced density matrix can be expressed in terms of the matrix $\Sigma$ as given in (\ref{ent37}), it is straightforward to work out the entanglement entropy in terms of $\Sigma$ as well.
To simplify calculations later down the line we will find the von Neumann entanglement entropy through a R\'enyi-like entropy. For this purpose let us define
\beq
\tilde{S}_q \equiv \Tr \rho_{\rm red}^q
\label{ent52}
\eeq
The entanglement entropy is then given by
\beq
S_{\rm ent} = -  \Tr \, \rho_{\rm red} \, \log \, \rho_{\rm red} = - \partial_q \tilde{S}_q |_{q=1}
\label{ent53}
\eeq
Using the result in (\ref{ent37}),
\beq
\tilde{S}_q = \sum_{n_1 = 0}^{\infty}\cdots (\det{(1-\Lambda)})^q  \prod_i  (\Lambda_i^q)^{n_i} = \det \left( {(1-\Lambda)^q \over 1-\Lambda^q} \right)
\label{ent54}
\eeq
It is now straightforward to find the von Neumann entanglement entropy explicitly as
\beq
S_{\rm ent} =  - \Tr \left( \log (1 - \Lambda) + {\Lambda \over 1-\Lambda} \log \Lambda \right)
\label{ent55}
\eeq
where, from (\ref{ent37}), $\Lambda = 2 \Sigma / (1+\sqrt{1-2\Sigma})^2$.
%%%%%%%%%%%%%%%%%%%%%%%%%%%%%%%%%%%%%%%
%%%%%%%%%%%%%%%%%%%%%%%%%%%%%%%%%%%%%%%
\subsection{Direct calculation of $\Sigma$}
%%%%%%%%%%%%%%%%%%%%%%%%%%%%%%%%%%%%%%%
%%%%%%%%%%%%%%%%%%%%%%%%%%%%%%%%%%%%%%%
Following equation (\ref{ent40}) we can write $\Sigma = A^T A$, where $A^T = {1\over \sqrt{2}}( \Omega_{\rm L} )^{-\half}\, \Omega_{\rm LR} \, (\Omega_{\rm R})^{-\half}$, and $\Omega_{\rm L}$, $\Omega_{\rm R}$ and $\Omega_{\rm LR}$ are defined in (\ref{ent27}). 
First, call the Dirichlet eigenvalues $(\omega_L)_a = a \pi / b$ and $(\omega_R)_\a = \a \pi /(L-b)$. (These are different from the matrices 
$(\Omega_{\rm L})_{ab}$ and $(\Omega_{\rm R})_{\a\b}$\,). Using (\ref{ent7}) and (\ref{ent20}) in (\ref{ent21}),
\beqar
(\Omega_{\rm L})_{ac} &=& {2 \over b} (-1)^{a+c} (\omega_{\rm L})_a (\omega_{\rm L})_c \, \sum_N \, {F_N^2 \, \omega_N \, \sin^2(\omega_N b) \sin^2 (\omega_N (L-b) ) \over (\omega_N^2 - (\omega_{\rm L})_a^2)(\omega_N^2-(\omega_{\rm L})_c^2)} \nonumber\\
(\Omega_{\rm R})_{\a \b} &=& {2 \over (L- b)} (-1)^{\a+\b} (\omega_{\rm R})_\a (\omega_{\rm R})_\b\, \sum_N \, {F_N^2 \, \omega_N \, \sin^2(\omega_N b) \sin^2 (\omega_N (L-b) )  \over (\omega_N^2 - (\omega_{\rm R})_\a^2)(\omega_N^2-(\omega_{\rm R})_{\b}^2)} \label{ent56} \\
(\Omega_{\rm LR})_{a \a} &=& {2 \over \sqrt{b(L-b)}} (-1)^{a+\a} (\omega_{\rm L})_a (\omega_{\rm R})_\a \, \sum_N \, {F_N^2 \, \omega_N \, \sin^2(\omega_N b) \sin^2 (\omega_N (L-b) )  \over (\omega_N^2 - (\omega_{\rm L})_a^2)(\omega_N^2-(\omega_{\rm R})_\a^2)} \nonumber
\eeqar
where $F_N$ is given in (\ref{ent9}), and $\{ \omega_N \}$ are solutions to the eigenvalue equation in (\ref{ent8}).

It is useful to rewrite these matrices in terms of integrals by noticing that $ 1/ (z \sinh z L + \kappa \sinh z b \, \sinh z (L-b) )$ (which is the function on the left side of the eigenvalue equation (\ref{ent8}) if we set $z = i \omega$) has
poles at $z =\pm i \omega_N$ with residues given by $ \pm i F_N^2 \sin (\omega_N b) \sin (\omega_N (L-b)) / 2 \omega_N$.
We now start by defining an integral
\beqar
\left(\Delta_{\rm L}\right)_{a c} &=& -{2 \over b} (-1)^{a+c}(\omega_{\rm L})_a (\omega_{\rm L})_c\, \nonumber\\
&& \hskip .2in \int_{-\infty}^{\infty} {dx \over \pi} { \, x^2 \sinh x b \sinh x(L-b) \over (x^2 + (\omega_{\rm L})_a^2)(x^2+(\omega_{\rm L})_c^2) (x \sinh xL + \kappa \sinh xb \sinh x(L-b))}\nonumber\\
 &=&- {2 \over  b}(-1)^{a+c} (\omega_{\rm L})_a (\omega_{\rm L})_c  \, \label{ent57}\\
&& \hskip .2in \oint_\Gamma \, {dz \over \pi} \, {z^2 \, \sinh zb \sinh z(L-b) \over (z^2 +(\omega_{\rm L})_a^2)(z^2 + (\omega_{\rm L})_c^2)(z \sinh zL +\kappa \sinh zb \sinh z(L-b))} \nonumber\\
&=&  (\Omega_{\rm L})_{ac} - \delta_{ac} (\omega_{\rm L})_a  \nonumber
\eeqar
where $\Gamma$ is the contour of a large semicircle in the upper 
half plane. Evaluating the integral by use of residues, we can relate this integral to $\Omega_{\rm L}$ as in the last line of this equation.
Hence, we see that we can write the matrix $(\Omega_{\rm L})_{ac}$ as
\beq
(\Omega_{\rm L})_{ac} = (\omega_{\rm L})_a \, \delta_{ac}+ (\Delta_{\rm L})_{ac}
\label{ent58}
\eeq
One can rewrite both $(\Omega_{\rm R})_{\a \b}$ and $(\Omega_{\rm LR})_{a \a}$ in an analogous way
\beqar
(\Omega_{\rm R})_{\a\b} &=& (\omega_{\rm R})_{\a} \, \delta_{\a\b} + (\Delta_{\rm R})_{\a\b} \nonumber\\
(\Omega_{\rm LR})_{a \a} &=& (\Delta_{\rm LR})_{a \a}
\label{ent59}
\eeqar
where
\beqar
(\Delta_{\rm L})_{a c} &=& - {4 \over b} (-1)^{a +c} \int_{0}^{\infty} {dx \over \pi} \, { (\omega_{\rm L})_a \over x^2 + (\omega_{\rm L})_{a}^2} \,{ (\omega_{\rm L})_c \over x^2 + (\omega_{\rm L})_{c}^2} 
\, {x^2 \over f(x, \kappa)}\nonumber\\
(\Delta_{\rm R})_{\a \b} &=& - {4 \over L-b} (-1)^{\a +\b} \int_{0}^{\infty} {dx \over \pi} \, { (\omega_{\rm R})_\a \over x^2 + (\omega_{\rm R})_{\a}^2} \,{ (\omega_{\rm R})_\b \over x^2 + (\omega_{\rm R})_{\b}^2} 
\, {x^2 \over f(x, \kappa)} \label{ent60}\\
(\Delta_{\rm LR})_{a \a} &=& - {4 \over \sqrt{b(L-b)}} (-1)^{a +\a} \int_{0}^{\infty} {dx \over \pi} \, { (\omega_{\rm L})_a \over x^2 + (\omega_{\rm L})_{a}^2} \,{ (\omega_{\rm R})_\a \over x^2 + (\omega_{\rm R})_{\a}^2} 
\,{x^2 \over f(x, \kappa)} \nonumber
\eeqar
and
\beq
f(x, \kappa) =  x \bigl[ \coth xb + \coth x (L-b) \bigr] + \kappa
\label{ent61}
\eeq
These integrals can be evaluated numerically. Moreover, one can also check numerically that a very good approximation is given by
\beq
f(x, \kappa) \approx 2x + \kappa
\label{ent62}
\eeq
This approximation is also justified in the large volume limit, $b, L-b \rightarrow \infty$. In this approximation the integrals can be carried out exactly to obtain
\beqar
\Delta( \omega, \omega') &\equiv& \int_0^{\infty} {dx \over \pi} {\omega \over x^2+\omega^2}\,{\omega' \over x^2 + \omega'^2}\,{x^2 \over 2x +\kappa} \nonumber\\
&=&{ \tilde{\omega}  \tilde{\omega}' \over 2 \pi} \left[{1\over  \tilde{\omega}^2 -  \tilde{\omega}'^2}\left( { { \tilde{\omega}^2 \over 1+  \tilde{\omega}^2} \log  \tilde{\omega} - { \tilde{\omega}'^2 \over 1 +  \tilde{\omega}'^2} \log \tilde{\omega}' }\right) \right.\nonumber\\
&&\hskip .3in \left.- {\pi \over 2} { \tilde{\omega}  \tilde{\omega}'-1 \over ( \tilde{\omega}+ \tilde{\omega}')(1+ \tilde{\omega}^2)(1+ \tilde{\omega}'^2)}\right]
\label{ent63}
\eeqar
where $\tilde{\omega} \equiv {2 \omega \over \kappa}$.
This function decreases monotonously as $\kappa$ increases. In the limit of large $\kappa$, $\kappa \rightarrow \infty$, $\Delta ( \tilde{\omega}, \tilde{\omega}') \sim 1/\kappa \rightarrow 0$. In this case we clearly reproduce the Dirichlet boundary condition at the interface, with the results
$(\Omega_{\rm L})_{ab} = (\omega_{\rm L})_a \delta_{ab}$, $(\Omega_{\rm R})_{\a\b} = (\omega_{\rm R})_\a \delta_{\a\b}$ and $\Omega_{\rm LR} = 0$. Since $\Omega_{\rm LR} =0$, it follows that $\Sigma = 0$, and from (\ref{ent55}), that $S_{\rm ent} = 0$, i.e., there is no entanglement.

In fact, since $\Delta$ is a monotonously decreasing function of $\kappa$, $\Sigma$ will also decrease monotonously as $\kappa$ increases. Therefore, as expected, as the interface approaches that of a perfectly conducting plate, the amount of entanglement between the left and right sides decreases.

For small $\kappa$ (or $\tilde{\omega} \gg 1$),
\beq
\Delta( \omega,  \omega') = {\omega \omega' \over 2 \pi} \left( {\log \omega /\omega' \over \omega^2-\omega'^2} - {\pi \kappa \over 4}{1 \over (\omega + \omega')\omega \omega'} + \O(\kappa^2) \right)
\label{ent64}
\eeq
The maximum value of $\Delta( \omega,  \omega')$ is along the diagonal,
i.e., for $\omega = \omega'$, with 
$\Delta( \omega,  \omega) \rightarrow 1/(4\pi)$ as $\omega \rightarrow
\infty$. 
Therefore, the behavior of
$\Delta (\omega,\omega')$ is no worse than
$\sim 1$ for large values of $\omega, \, \omega'$. Consequently, $(\Delta_{\rm L})_{ab} \ll (\omega_{\rm L})_a$ and $(\Delta_{\rm R})_{\a\b} \ll (\omega_{\rm R})_{\a}$ in the large matrix limit, $a, \a \gg 1$. (In fact, it is easy to check that even for small values of $a$ and $\a$, the Dirichlet part dominates.) 
We can, therefore, make the approximation 
\beq
(\Omega_{\rm L})_{ab} \approx (\omega_{\rm L})_a \, \delta_{ab}, \hskip .3in
(\Omega_{\rm R})_{\a\b} \approx (\omega_{\rm R})_\a \, \delta_{\a\b}
\label{ent65}
\eeq
With these values, $A^T_{a\a}$ as defined in (\ref{ent40})
can be approximated as
\beq
A^T_{a \a} \approx {(\Delta_{\rm LR})_{a\a} \over \sqrt{2  (\omega_{\rm L})_a \, (\omega_{\rm R})_{\a} }} =  -\sqrt{ 8 \over b(L-b)} \,  (-1)^{a+\a} \,  {\Delta ( \omega_{{\rm L}a},\omega_{{\rm R} \a} ) \over \sqrt{ (\omega_{\rm L})_a (\omega_{\rm R})_\a}}
\label{ent66}
\eeq
and
\beq
\Sigma_{ac} = {8 \over b(L-b)} \, {(-1)^{a+c} \over \sqrt{ (\omega_{\rm L})_a (\omega_{\rm L})_c}} \sum_{\a =1}^{\infty} {\Delta(\omega_{{\rm L}a}, \omega_{{\rm R}\a}) \Delta( \omega_{{\rm R} \a}, \omega_{{\rm L}c}) \over (\omega_{\rm R})_\a}
\label{ent67}
\eeq
For large values of $b$ and $L-b$, the summation in this expression can be approximated by an integral along the lines of the Euler-Maclaurin formula.
We can then find an expression for $\Sigma$ for small $\kappa$
using $\Delta (\omega, \omega')$ given in (\ref{ent64}).
This is given by
\beqar
\Sigma_{ac} &=& \Sigma_{ac}(\kappa = 0 ) + \kappa \,\delta \Sigma_{ac}
\nonumber\\
\Sigma_{ac}(\kappa = 0)&=&{1\over 3 \pi^4} (-1)^{a+c} \sqrt{ac} \left( {\pi^2 \log(a/c) + \log^3(a/c) \over a^2 -c^2} + \O\left( {\log a \log c \over a^2 c^2} \right) \right)
\label{ent67a}\\
\delta \Sigma_{ac} &=& -{b \over 4 \pi^4} (-1)^{a+c} {1 \over \sqrt{ac}} \left({\pi^2 + \log^2(a/c) \over a+c} + \O\left({ \log a +\log c \over ac} \right) \right)
\label{ent67b}
\eeqar

%%%%%%%%%%%%%%%%%%%%%%%%%%%%%%%%%%%%%%%
%%%%%%%%%%%%%%%%%%%%%%%%%%%%%%%%%%%%%%%
\subsection{Entropy for various $\kappa$}
%%%%%%%%%%%%%%%%%%%%%%%%%%%%%%%%%%%%%%%
%%%%%%%%%%%%%%%%%%%%%%%%%%%%%%%%%%%%%%%
As mentioned in subsection \ref{entropy} to find the entanglement entropy it is easiest to work with a R\'enyi-like entropy defined in (\ref{ent52}). For this we need traces over powers of $\Sigma$, i.e., $\Tr \Sigma^N$. 
Looking back at (\ref{ent67}), we obtain
\beq
\Sigma^N_{ac} = \left({8\over b(L-b)}\right)^N { (-1)^{a+c} \over \sqrt{ (\omega_{\rm L})_a (\omega_{\rm L})_c}} \, \sum_{\a_1} \sum_{a_1} \cdots \sum_{a_{N-1}} \sum_{\a_N} {\Delta(a,\a_1)\Delta(\a_1,a_1) \cdots \Delta(\a_N, c) \over (\omega_{\rm R})_{ \a_1} (\omega_{\rm L})_{ a_1} \cdots (\omega_{\rm R})_{\a_N} }
\label{ent68}
\eeq
where $\Delta(a, \a) \equiv \Delta( \omega_{{\rm L}a}, \omega_{{\rm R}\a} )$. 
In calculating $\Tr\, \Sigma^N$ we take the large $b$ and $L-b$ limit, thus, approximating sums with integrals by taking $\sum_{a=1}^{\infty} \rightarrow \int_0^{\infty} da$. This leads to
\beq
\Sigma^N_{ac} = \left( {8 \over \pi^2} \right)^N {(-1)^{a+c} \over \sqrt{ac}} \int_0^{\infty} {d\omega_1 \over \omega_1} \cdots {d\omega_{2N-1} \over \omega_{2N-1}} \Delta(\omega_{{\rm L}a}, \omega_1) \Delta(\omega_1, \omega_2) \cdots \Delta(\omega_{2N-1},\omega_{{\rm L}c})
\label{ent69}
\eeq
The dominant part in $\Tr\,\Sigma^N$ comes from the ultraviolet (UV) limit of the trace, i.e., for $a, c \gg 1$. In fact, $\Tr\, \Sigma^N$ is UV divergent for any finite $\kappa$.  
The leading order UV term comes from the $\kappa$-independent part of $\Delta(\omega, \omega')$ in (\ref{ent64}). Using it in (\ref{ent69}),
and performing appropriate coordinate transformations, we get
\beqar
\Sigma^N_{ac}(\kappa=0) &=& \left({2 \over \pi^4}\right)^N (-1)^{a+c} \sqrt{ac}  \int_0^{\infty} dy_1 y_1 \cdots dy_{2N-1} y_{2N-1} { \log(a/y_1) \over a^2-y_1^2} \cdots {\log(y_{2N-1}/c) \over  y_{2N-1}^2 - c^2} \nonumber\\
&=& \left({2\over \pi^4}\right)^N  (-1)^{a+c} \sqrt{ac} \int_0^{\infty} dx_1 x_1 \cdots dx_{2N-1} x_{2N-1} {\log x_1 \over x_1^2 -1}  \cdots { \log (x_{2N-1})\over x_{2N-1}^2 -1} \nonumber\\
&& \hskip 2in \times {\log ({a \over c} x_1 \cdots x_{2N-1}) \over (a x_1 \cdots x_{2N-1})^2 - c^2}
\label{ent70}
\eeqar
In going to the second line of this equation, we have made
the change of variables $y_1 = a x_1$, $y_2 = x_2 y_1 = a x_2 x_1 $, etc.
This integral can be evaluated exactly using Mathematica as
\beq
\Sigma^N_{ac}(\kappa=0) = {1\over \Gamma(4N)}\left({2 \over \pi^4} \right)^N (-1)^{a+c} \sqrt{ac}\,  {\log (a/c) \over a^2-c^2} \prod_{n=1}^{2N-1} (\log^2(a/c) + n^2 \pi^2)
\label{ent71}
\eeq 
where $\Gamma(4N) = (4N-1)!$ is Euler's Gamma function.

The trace of $\Sigma^N$ is, therefore, of the form
\beq
\Tr \Sigma^N = \int_0^{\infty} da \, \Sigma^N_{aa} = {2^N \over 2\pi^2} {\Gamma(2N)^2 \over \Gamma(4N)} \, \int_0^{\infty} {da \over a} +\O(\kappa)
\label{ent72}
\eeq
There is a logarthmic UV divergence and and infrared (IR) divergence in the integral over $a$. (Strictly speaking there is no IR divergence since the sum starts at $a, c = 1$. The IR divergence appears because we are taking the large $b$ and  $L-b$ approximation.) We can regularize the UV divergence by capping the size of $\Sigma_{ac}$, i.e., by taking $a ,c \leq \mathcal{N}$. 
For the IR divergence we take $a, c \geq 1$, in analogy to $\omega_{\rm L} \geq \pi/b$. This leads to
\beq
\Tr \Sigma^N = {2^N \over 2\pi^2} {\Gamma(2N)^2 \over \Gamma(4N)} \, \log \mathcal{N} + \O(\kappa)
\label{ent73}
\eeq

For working out the entanglement entropy it is useful to notice that the matrix elements $\Sigma_{ac}$ are very small,
being of the order ${1 \over 6 \pi^2} \sim 10^{-2}$ for $a,c \sim 1$ and decreasing
rapidly for higher values of $a,\, c$, so that
it is a good working approximation to take $\Sigma_{ac} \ll 1$.
Therefore, we can approximate $\Lambda$, as defined in (\ref{ent37}),
by $\Lambda \simeq \Sigma /2$.
Correspondingly, for $\tilde{S}_q$,
defined in (\ref{ent54}), we can write
\beq
\tilde{S}_q = \det \left({(1-\Lambda)^q \over 1 -\Lambda^q}\right) \simeq e^{- \Tr ( q \Sigma /2 - (\Sigma/2)^q )}   \sim e^{-(q\,f(1)-f(q)) \log \mathcal{N}}
\label{ent74}
\eeq
where 
\beq
f(N) =  {1 \over 2\pi^2} {\Gamma(2N)^2 \over \Gamma(4N)}
\label{ent75}
\eeq
The leading order UV divergent term in the von Neumann entropy
$S_{\rm ent}$ is hence
\beq
S_{\rm ent} = - \partial_q \tilde{S}_q \Big|_{q=1} \sim  (f(1) -f'(1)) \log \mathcal{N}
\approx (0.0366)\,\log\mathcal{N}
\label{ent76}
\eeq
The logarithmic divergence is, of course, in accordance with general
expectations. For a field theory in $(d+1)$-dimensional spacetime,
the divergent term of the entropy should behave as the area of the interface,
so that $S_{\rm ent} \sim {\rm Area}\int^{1/\e}
{d^dk /\vert k\vert}$, where $\e$ is a short-distance cutoff. For us, since we are in 1+1 dimensions, the 
expected divergence (corresponding to the $d\rightarrow 1$ limit)
should be logarithmic, as in (\ref{ent76}).

\begin{figure}[!b]
\begin{center}
\begin{tikzpicture}
\pgftext{
\scalebox{.6}{\includegraphics{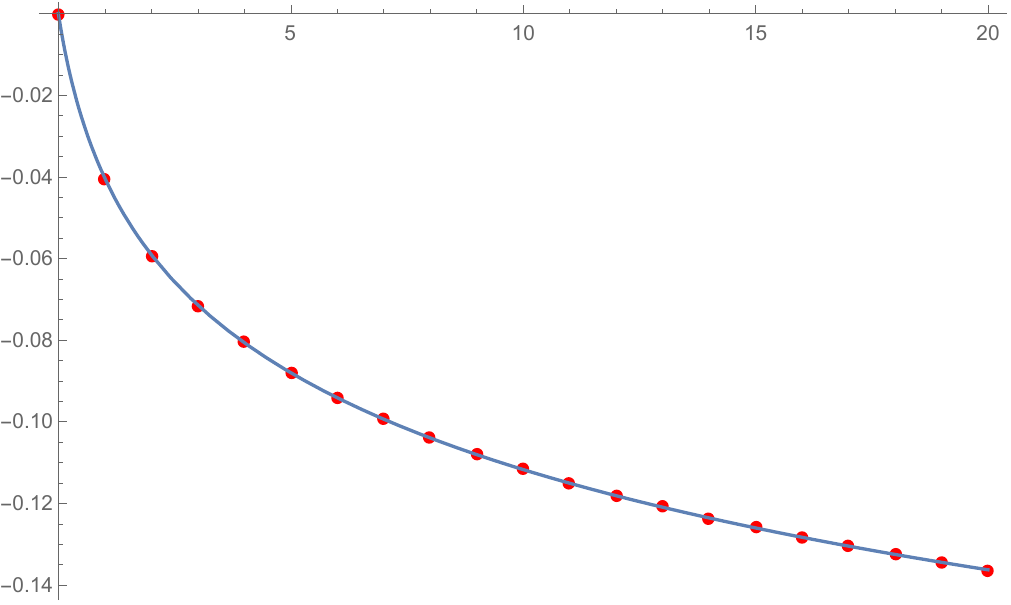}}
}
\draw(1,+3.4)node{$\kappa b / 2\pi$};
\draw(-6,.8)node{$S_{\rm ent}^{\rm fin}(\kappa)$};
\end{tikzpicture}
\caption{The red points are values of $S_{\rm ent}^{\rm fin}(\kappa)$ for ${\kappa b \over2\pi} = 0$ to $20$. The blue line is a logarithmic fit given in (\ref{ent76d}).}
\label{Fig1a}
\end{center}
\end{figure}

The dependence of $S_{\rm ent}$ on $\kappa$ will in general be complicated. 
We can evaluate it numerically by using (\ref{ent69}) with the cutoffs introduced above,
\beqar
\Tr \Sigma^N &=& \left({8 \over \pi^2} \right)^N \int_1^{\mathcal{N}} {da \over a} \int_0^{\infty} {d\omega_1 \over \omega_1} \cdots {d\omega_{2N-1} \over \omega_{2N-1}} \Delta(\omega_{{\rm L} a}, \omega_1) \cdots \Delta(\omega_{2N-1}, \omega_{{\rm L} a}) \nonumber\\
&=& \Tr \Sigma^N_{\rm UV} + \Tr \Sigma^N_{\rm fin}
\label{ent76a}
\eeqar
where $\Tr\Sigma^N_{\rm UV}$ is the UV-divergent part in (\ref{ent73}) and $\Tr \Sigma^N_{\rm fin}$ has the $\kappa$-dependence. The entanglement entropy will then be
\beq
S_{\rm ent} (\kappa) = S_{\rm ent}^{\rm UV} + S_{\rm ent}^{\rm fin}(\kappa)
\label{ent76b}
\eeq
where $S_{\rm ent}^{\rm UV} \sim \log \mathcal{N}$ as in (\ref{ent76}). 
In order to approximate $S_{\rm ent}^{\rm fin}$ numerically, we find that for large enough $\kappa$ ($\kappa b /2\pi \gtrsim 1$) a good fit for $\Tr \Sigma^N_{\rm fin}$ is given by analogy to (\ref{ent73}),
\beq
\Tr\Sigma^N_{\rm fin} = - 2^N \,a\, {\Gamma (2b N)^2 \over \Gamma (4c N)}, \hskip.2in a= a(\kappa), \, b= b(\kappa) , \,c= c(\kappa)
\label{ent76c}
\eeq
Using this fit we can numerically approximate $\partial_N \Tr \Sigma^N_{\rm fin} \big|_{N=1}$ and, hence, $S_{\rm ent}^{\rm fin}$, for various values of $\kappa$.
In Fig.\,\ref{Fig1a} below we plot $S_{\rm ent}^{\rm fin} (\kappa)$ for ${\kappa b \over 2\pi} = 0$ to $20$, together with a logarithmic fit.

In the large $\kappa$ limit, $S_{\rm ent}^{\rm fin} (\kappa)$ is numerically well approximated by the logarithmic function
\beq
S_{\rm ent}^{\rm fin} (\kappa) = - A \log \left(1 + {\kappa b \over \pi }\right), \hskip .2in A= 0.0367
\label{ent76d}
\eeq
This logarithmic dependence of the entropy for large $\kappa$ is as expected.
Since the states are characterized by the eigenvalues of the Laplacian,
in the large $b$, $L-b$ limit, the leading term in the entropy should only
be a function of the cutoff and $\kappa$.
Therefore, given that the UV divergence is logarithmic,  and
converting the cutoff into the proper dimensionful units for the eigenvalue
as $\omega_{\rm max} = {\mathcal N} \pi /b$,
we should expect $S_{\rm ent} \sim A \log[\omega_{\rm max}/\kappa]$,
with the same coefficient $A$ for both $\log\omega_{\rm max}$
and $\log \kappa$.
This is exactly what we find; notice that the coefficient of $\log{\mathcal N}$
in (\ref{ent76}) and $A$ in (\ref{ent76d}) are the same to within the numerical
precision we have.
In fact, combining (\ref{ent76d})
with the UV divergent part in (\ref{ent76}), the large $\kappa$ behavior 
can be expressed as
\beq
S_{\rm ent} (\kappa) \sim (0.037) \log \left(\mathcal{N} \pi / \kappa b \right)
\label{ent76d2}
\eeq

In the small $\kappa$ limit the logarithmic fit diverges slightly, but we numerically find the first order dependence on $\kappa$ to be
\beq
S_{\rm ent}^{\rm fin} (\kappa) \simeq - (0.099) {\kappa b \over 2 \pi} + \O(\kappa^2)
\label{ent76e}
\eeq

%%%%%%%%%%%%%%%%%%%%%%%%%%%%%%%%%%%%%%%
%%%%%%%%%%%%%%%%%%%%%%%%%%%%%%%%%%%%%%%

The first order term in the small $\kappa$ limit can be found analytically using (\ref{ent64}) and (\ref{ent69}). 
\beq
\delta (\Tr \Sigma^N) = 2N \left({8 \over \pi^2} \right)^N \int_0^{\infty} {d \omega_1 \over \omega_1} \cdots {d\omega_{2N} \over \omega_{2N}} \Delta (\omega_1, \omega_2) \cdots \Delta (\omega_{2N-2}, \omega_{2N-1}) \delta \Delta (\omega_{2N-1}, \omega_1)
\label{ent77}
\eeq
where $\Delta$ is the $\kappa$-independent term and $\delta \Delta \sim \kappa$ is the first order term in $\kappa$ in (\ref{ent64}).
Performing similar coordinate transformations as in (\ref{ent70}),
\beqar
\delta (\Tr \Sigma^N) &=& -{\pi \kappa \over 2} N C_N \int_0^{\infty} { d\omega_1 \over \omega_1^2} \nonumber\\
&\rightarrow&  -{\pi \kappa \over 2} N C_N \int_{\pi/b}^{\infty} { d\omega_1 \over \omega_1^2} = - {\kappa b \over 2} N C_N
\label{ent78}
\eeqar
In going to the second line we included the IR cutoff on $\omega_{\rm L} \leq \pi/b$. The factor $C_N$ is given by the integral
\beq
C_N = \left({2\over \pi^4}\right)^N  \int_0^{\infty} dx_1 \cdots dx_{2N-1}{\log x_1 \over x_1^2 -1}  \cdots { \log x_{2N-1} \over x_{2N-1}^2 -1} {1 \over x_1 \cdots x_{2N-1} +1}
\label{ent79}
\eeq
These integrations can be done recursively by noticing that
\beq
\int_0^{\infty} dx {\log x \over x^2-1}{1 \over ax+1} = {\pi^2 \over 4} {1 \over a+1}+ {a \log^2 a \over 2(a^2-1)}
\label{ent80}
\eeq
The integration over $x_{2N-1}$ leads to the result on the right hand side
with $a= x_1 \cdots x_{2N-2}$. The first term on the right hand side thus
leads to an integral over $x_{2N-2}$ which is of the same form and the process can be continued.
As for the second term on the right hand side of
(\ref{ent80}), it will produce more and
more even powers of $\log$'s when we integrate over
$x_{2N-2}$, $x_{2N-3}$, etc.
The final result will vanish since
for the final integration we have $a =1$.
The first term produces powers of $\pi^2/4$ for
$2N -2$ integrations and $\pi^2/8$ for the last one. The final result is thus
\beq
C_N = {2 \over \pi^2 8^N}, \hskip .3in
\delta (\Tr \Sigma^N)  = - {\kappa b  \over \pi^2} { N \over 2^{3N}}
\label{ent81}
\eeq

The first $\kappa$-dependent correction (for small
$\kappa$) to the R\'enyi entropy in (\ref{ent54}) can now be calculated as
\beqar
\tilde{S}_q  &\simeq& \tilde{S}_q (\kappa =0 ) \, \exp\left( - {\kappa b} g(q) + \O(\kappa^2) \right)\nonumber\\
g(q) &\simeq& -{ q \over \pi^2} \left(1/16 - 1/16^q\right)
\label{ent83}
\eeqar
Using this, the first $\kappa$-dependent correction to the von Neumann
entanglement entropy is hence obtained as
\beq
S_{\rm ent}^{\rm fin} \simeq g'(1)\, \kappa b + \O(\kappa^2) = - {\log2 \over 4 \pi^2} \kappa b + \O(\kappa^2) \simeq - (0.1103) {\kappa b \over 2 \pi} +\O(\kappa^2)
\label{ent85}
\eeq
which is in good agreement with the numerical value in (\ref{ent76e}).
%%%%%%%%%%%%%%%%%%%%%%%%%%%%%%%%%%%%%%%
%%%%%%%%%%%%%%%%%%%%%%%%%%%%%%%%%%%%%%%
\section{Correlations and the reduced density matrix}
%%%%%%%%%%%%%%%%%%%%%%%%%%%%%%%%%%%%%%%
%%%%%%%%%%%%%%%%%%%%%%%%%%%%%%%%%%%%%%%
The existence of correlations over spacelike separations is the main characteristic of entanglement in relativistic field theories.
It is therefore useful to relate the reduced density matrix 
to correlation functions. 
%%%%%%%%%%%%%%%%%%%%%%%%%%%%%%%%%%%%%%%
%%%%%%%%%%%%%%%%%%%%%%%%%%%%%%%%%%%%%%%
\subsection{Relating $\rho_{\rm red}$ to correlations}
%%%%%%%%%%%%%%%%%%%%%%%%%%%%%%%%%%%%%%%
%%%%%%%%%%%%%%%%%%%%%%%%%%%%%%%%%%%%%%%
Starting with the expression for
$\rho_{\rm red}$ given in
(\ref{ent28}), we can calculate various correlations functions.
When both
arguments $x$ and $y$ of the propagator
are taken to be in the left region, we can write
\beqar
(G_{\rm LL})_{ab} &=& \int_{x,y} v_a (x) \, v_b (y) \la \vf (x) \vf (y)\ra
= \la b_a b_b\ra\nonumber\\
&=& \int [db d{\tilde b}] \, \rho_{\rm red} (b, b) \, b_a b_b
\nonumber\\
&=& {1\over 2} \left( \Omega_{\rm L}  - \Omega_{\rm LR} \Omega_{\rm R}^{-1} \Omega_{\rm RL} \right)^{-1}_{ab}
\label{cor1}
\eeqar
where the subscript ${\rm LL}$ on $G$ indicates that both points are in the left region.
In a similar way, we can write
\beqar
(G_{\rm RR})_{\alpha \beta}&=& {1\over 2} \left( \Omega_{\rm R} 
- \Omega_{\rm RL} \Omega_{\rm L}^{-1} \Omega_{\rm LR} \right)^{-1}_{\alpha\beta}\nonumber\\
(G_{\rm LR})_{a\alpha}&=& - {1\over 2} \left[
\Omega^{-1}_{\rm L} \Omega_{\rm LR} \left( \Omega_{\rm R} 
- \Omega_{\rm RL} \Omega^{-1}_{\rm L} \Omega_{\rm LR} \right)^{-1}\right]_{a\alpha}\label{cor2}\\
&=& - \left[ \Omega^{-1}_{\rm L} \Omega_{\rm LR} G_{\rm RR} \right]_{a\alpha}
\nonumber\\
(G_{\rm RL})_{\alpha a} &=&  - \left[ \Omega^{-1}_{\rm R} \Omega_{\rm RL} G_{\rm LL} \right]_{\alpha a}\nonumber
\eeqar
Since $G_{\rm LL}$, $G_{\rm RR}$ are symmetric and
$G_{\rm LR} = G_{\rm RL}^T$, we can also write
\beq
(G_{\rm LR})_{a\a} = - (G_{\rm LL} \Omega_{\rm LR} \Omega_{\rm R}^{-1} )_{a\a}, \hskip .3in
(G_{\rm RL})_{\a a} =
- (G_{\rm RR} \Omega_{\rm RL} \Omega_{\rm L}^{-1})_{\a a}
\label{cor2a}
\eeq
Equivalently, these relations (\ref{cor2}), (\ref{cor2a}) can also be written as 
\beqar
\left(\Omega_{\rm L} G_{\rm LL} + \Omega_{\rm LR} G_{\rm RL}\right)_{ab} &=& {1\over 2}
\delta_{ab}\nonumber\\
\left(\Omega_{\rm R} G_{\rm RR} + \Omega_{\rm RL} G_{\rm LR}\right)_{\a \b} &=& {1\over 2}
\delta_{\a \b}
\label{cor2b}\\
G_{\rm LR} \Omega_{\rm R} + G_{\rm LL} \Omega_{\rm LR} 
&=& 0 \nonumber\\
G_{\rm RL} \Omega_{\rm L} + G_{\rm RR} \Omega_{\rm RL} &=& 0
\nonumber
\eeqar
We can invert these equations and express $\Omega_{\rm L}$,
$\Omega_{\rm R}$ and $\Omega_{\rm LR}$, and hence the reduced density matrix, in terms of the
correlations restricted to the left-left, right-right, left-right regions.
We find
\beq
G_{\rm LR} G^{-1}_{\rm RR} G_{\rm RL} G^{-1}_{\rm LL} =
\Omega_{\rm L}^{-1} \Omega_{\rm LR} \Omega_{\rm R}^{-1} \Omega_{\rm RL}
\label{cor3}
\eeq
Using this relation, and (\ref{cor1}), we can write
\beq
{1\over 2} \Omega_{\rm L}^{-1} = 
G_{\rm LL} - G_{\rm LR} G^{-1}_{\rm RR} G_{\rm RL}
\label{cor4}
\eeq
This allows for the calculation of $\Omega_{\rm L}$ directly in terms of the correlations.
Further, we can rewrite $\Sigma$ as
\beqar
\Sigma &=& {1\over 2} \Omega_{\rm L}^{-{\half}}
( \Omega_{\rm LR} \Omega_{\rm R}^{-1} \Omega_{\rm RL} ) \Omega_{\rm L}^{-{\half}}\nonumber\\
&=&{1\over 2} \Omega_{\rm L}^{{\half}} \,
(G_{\rm LR} G^{-1}_{\rm RR}  G_{\rm RL} )\, G^{-1}_{\rm LL}\, \Omega_{\rm L}^{-{\half}}\nonumber\\
&=&\Omega_{\rm L}^{{\half}}\, \mathbb{X} ( 1- G^{-1}_{\rm LL} \mathbb{X} )
\Omega_{\rm L}^{{\half}}
\label{cor5}\\
\mathbb{X}&=& G_{\rm LR} G^{-1}_{\rm RR} G_{\rm RL}
\label{cor6}
\eeqar
Equation (\ref{cor4}) can be used to calculate $\Omega_{\rm L}$
directly in terms of the correlation functions.
Likewise, $\mathbb{X}$ is given by (\ref{cor6}), which leads, in turn, to
$\Sigma$. The reduced density matrix $\rho_{\rm red}$ can thus be
constructed from the correlation functions using (\ref{cor4})-(\ref{cor6}).
This can provide a workable alternative route to $\rho_{\rm red}$
if we have a method of calculating the correlation functions
including effects due to the interface conditions on the interface.
We turn to this question now.
%%%%%%%%%%%%%%%%%%%%%%%%%%%%%%%%%%%%%%%
%%%%%%%%%%%%%%%%%%%%%%%%%%%%%%%%%%%%%%%
\subsection{Direct calculation of correlations}
%%%%%%%%%%%%%%%%%%%%%%%%%%%%%%%%%%%%%%%
%%%%%%%%%%%%%%%%%%%%%%%%%%%%%%%%%%%%%%%
The equal-time correlations can be calculated using the 
vacuum wave function directly. 
The correlation for the whole space is given by
\beq
\la \vf(x) \vf (y) \ra = \sum_N {u_{N}(x) u_N (y) \over 2 \omega_N}
\label{cor6a}
\eeq
For the left-left correlation function, we can then write
\beqar
(G_{\rm LL})_{ab}&=& \sum_N {S_{Na} S_{N b} \over 2 \omega_N}
\nonumber\\
&=& {2 \over b} (-1)^{a+b} (\omega_{\rm L})_a (\omega_{\rm L})_b
\sum_N {1\over 2 \omega_N} {F_N^2 \sin^2(\omega_N b) \sin^2(\omega_N (L-b)) \over (\omega_N^2 - (\omega_{\rm L})_a^2) (\omega_N^2 - (\omega_{\rm L})_b^2)}
\label{cor6b}
\eeqar
We can convert the summation to an integral, as we did after equation
(\ref{ent56}). Towards this, consider
\beqar
({\tilde \Delta}_{\rm L} )_{ab} &=&
{1 \over b} (-1)^{a+b} (\omega_{\rm L})_a (\omega_{\rm L})_b
\int_{-\infty}^\infty {d x \over \pi} 
{ \sinh (b x) \sinh ((L-b) x) \over (x^2 +(\omega_{\rm L})_a^2)
(x^2+ (\omega_{\rm L})_b^2)}\nonumber\\ 
&&\hskip .2in\times{1 \over \bigl[ x \sinh(L x) + \kappa
\sinh (b x) \sinh ((L-b) x)\bigr] }\nonumber\\
&=&{1 \over b} (-1)^{a+b} (\omega_{\rm L})_a (\omega_{\rm L})_b
\left[ 2 \sum_N {1\over 2 \omega_N} {F_N^2 \sin^2(\omega_N b) \sin^2(\omega_N (L-b)) \over (\omega_N^2 - (\omega_{\rm L})_a^2) (\omega_N^2 - (\omega_{\rm L})_b^2)} - \delta_{ab}{b \over 2 (\omega_{\rm L})_a^3 }
\right]\nonumber\\
&=& (G_{\rm LL})_{ab} - {1\over 2 (\omega_{\rm L})_a}\delta_{ab} 
\label{cor6c}
\eeqar
This shows that we can write $G_{\rm LL}$ as
\beqar
(G_{\rm LL})_{ab} &=& \delta_{ab} {1\over 2 (\omega_{\rm L})_a}
+ ({\tilde \Delta}_{\rm L} )_{ab}\nonumber\\
({\tilde \Delta}_{\rm L} )_{ab} &=& 
{2 \over b} (-1)^{a+b} \int_0^\infty {d x\over \pi} {(\omega_{\rm L})_a
\over x^2 + (\omega_{\rm L})_a^2} {(\omega_{\rm L})_b
\over x^2 + (\omega_{\rm L})_b^2} \, {1\over f(x, \kappa )}
\label{cor6d}
\eeqar
In a similar way
\beqar
(G_{\rm RR})_{\a \b} &=& \delta_{\a \b} {1\over 2 (\omega_{\rm R})_\a}
+ ({\tilde \Delta}_{\rm R} )_{\a \b}\nonumber\\
({\tilde \Delta}_{\rm R} )_{\a \b} &=& 
{2 \over (L-b)} (-1)^{\a+ \b} \int_0^\infty {d x\over \pi} {(\omega_{\rm R})_\a
\over x^2 + (\omega_{\rm R})_\a^2} {(\omega_{\rm R})_\b
\over x^2 + (\omega_{\rm R})_\b^2} \, {1\over f(x, \kappa )}
\label{cor6e}\\
(G_{\rm LR})_{a \a} &=& ({\tilde \Delta}_{\rm LR} )_{a \a}\nonumber\\
&=&{2 \over \sqrt{b(L-b)}} (-1)^{a+ \a} \int_0^\infty {d x\over \pi} {(\omega_{\rm L})_a
\over x^2 + (\omega_{\rm L})_a^2} {(\omega_{\rm R})_\a
\over x^2 + (\omega_{\rm R})_\a^2} \, {1\over f(x, \kappa )}
\label{cor6f}
\eeqar
The use of these expressions (\ref{cor6d})-(\ref{cor6f}) along with
(\ref{cor4}) and (\ref{cor5}) will give the reduced density matrix
directly in terms of the correlations.
%%%%%%%%%%%%%%%%%%%%%%%%%%%%%%%%%%%%%%%
%%%%%%%%%%%%%%%%%%%%%%%%%%%%%%%%%%%%%%%
\section{An alternate measure of entanglement}
%%%%%%%%%%%%%%%%%%%%%%%%%%%%%%%%%%%%%%%
%%%%%%%%%%%%%%%%%%%%%%%%%%%%%%%%%%%%%%%
The essence of entanglement is the spread of the wave function
which leads to correlations over spacelike separations.
In the nonrelativistic context, this is vividly illustrated by the phenomenon
of tunneling through a potential barrier where, even with the decay of the wave function into the classically forbidden region, there is a nonzero value for it beyond the 
forbidden region; this in turn leads to a nonzero probability of detecting the
particle beyond the barrier.
More generally, including the relativistic case, we can formulate this as follows.
With the partitioning of space as before,
we introduce a particle in the left region. 
The probability of detecting this anywhere in the whole of space
is obviously one.
However, we can ask about the probability of the particle being detected in the right region. This will be zero in the absence of entanglement between the left and right regions. A nonzero probability will thus be a measure of entanglement.

More quantitatively, we can formulate this as follows.
With the vacuum wave function given as in (\ref{ent17}), 
the lowest excited state will be of the form
\beq
\Psi^{(1)} = \C \, \vf (x) \, \Psi_0
\label{altent1}
\eeq
where $\C$ is a normalization factor.
If we restrict to the left region, a wave function of the form given above leads to
\beq
\Psi^{(1)}_a \sim \int_x v_a (x) \vf (x) \, \Psi_0
\label{altent2}
\eeq
The normalization integral for this set of wave functions is
\beq
\int [d\vf] \, \Psi^{(1)*}_a \Psi^{(1)}_b = \int_{x,y} v_a(x) v_b (y) \int_\vf \Psi_0^* \, \Bigl[\vf (x) \vf (y) \Bigr] \Psi_0 = (G_{\rm LL})_{ab}
\label{altent3}
\eeq
Since $(G_{\rm LL})_{ab}$ is a symmetric matrix, writing it as
$(G_{\rm LL})_{ab} = N^T_{ac} N_{cb}$, the normalized states of
the form (\ref{altent2}) are given by
\beq
\braket{\vf|a} = (N^T)_{ab}^{-1}\, \int_x v_b (x) \vf (x) \, \Psi_0
\label{altent4}
\eeq
Similarly, if we restrict to the right region, we can define a normalized set
of wave functions of the form (\ref{altent1}) by
\beq
\braket{\vf|\a} = (M^T)_{\a \b}^{-1}\, \int_x w_\beta (x) \vf (x) \, \Psi_0
\label{altent5}
\eeq
where $M^T M = G_{\rm RR}$.
The overlap between these two functions is given by
\beqar
\braket{a|\a} &=& (N^T)_{ab}^{-1} (M^T)_{\a \b}^{-1}\,
\int_{x,y} v_b (x) w_\beta (y)\, \la \vf(x) \vf(y)\ra\nonumber\\
&=& (N^T)_{ab}^{-1}  \, (G_{\rm LR})_{b\b} M^{-1}_{\beta \a}
\label{altent6}
\eeqar
This gives the probability amplitude that if we introduce a particle in the state
$\ket{a}$ in the left region it can be observed in the right region in the
state $\ket{\a}$.
We are interested in the total probability that, if we introduce the particle in the left region as above, we can observe it in the right region in any state.
Evidently, it is given by
\beqar
\P (R| L; a) &\equiv&\sum_\alpha \vert \braket{a|\a}\vert^2 =
(N^T)_{ab}^{-1} (G_{\rm LR})_{b\b} M^{-1}_{\beta \a} (M^T)^{-1}_{\a \gamma}
(G_{\rm RL})_{\gamma c} N^{-1}_{ca}\nonumber\\
&=&(N^T)_{ab}^{-1} (G_{\rm LR} G^{-1}_{\rm RR} G_{\rm RL} )_{bc}
N^{-1}_{ca}
\label{altent7}
\eeqar
This is the conditional probability of observing a particle anywhere in any state in the right region, given that it is introduced 
into the left region in state $\ket{a}$. 
This conditional probability can be used as a measure of entanglement. We will investigate how this can change with changing values of $\kappa$.
%%%%%%%%%%%%%%%%%%%%%%%%%%%%%%%%%%%%%%%
\begin{figure}[!t]
\begin{center}
\begin{tikzpicture}
\pgftext{
\scalebox{.6}{\includegraphics{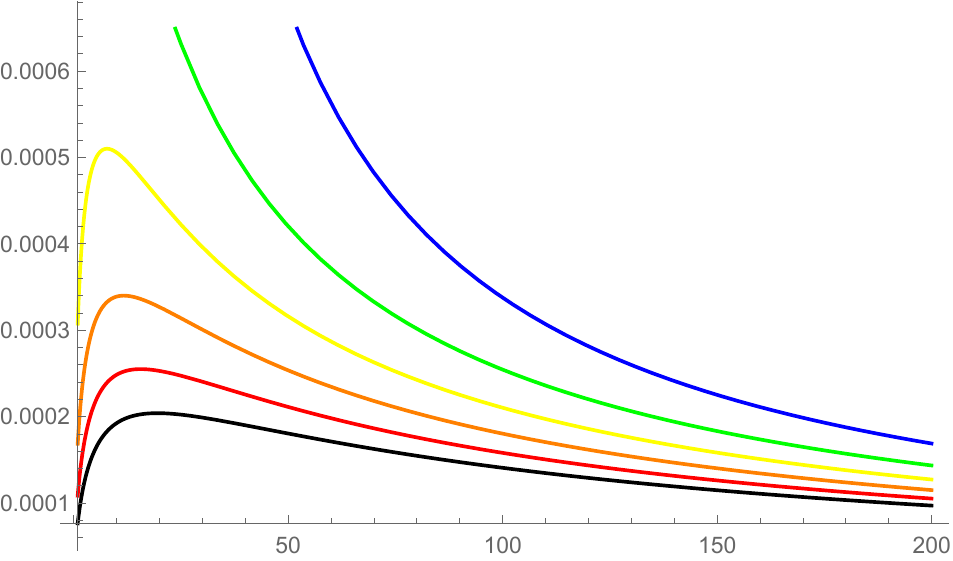}}
}
\draw(1,-3.2)node{$a \rightarrow$};
\draw(-5.7,.8)node{$\P(R|L; a)$};
\end{tikzpicture}
\caption{$\P(R|L; a)$ for $a = 1$ to $200$. The top curve is for
$(\kappa b/2\pi) = 0$, the lowest is for $(\kappa b/2\pi) = 50$.}
\label{Fig1}
\end{center}
\end{figure}
%%%%%%%%%%%%%%%%%%%%%%%%%%%%%%%%%%%%%%%

In the approximation of $G_{\rm LL} \approx (2 (\omega_{\rm L})_a)^{-1} \delta_{ab}$ as in (\ref{cor1}) combined with
(\ref{ent65}), we can simplify $\P(R|L; a)$ as
\beq
\P (R| L; a)  \approx 2 (\omega_{\rm L})^{{1\over 2}}_a 
(G_{\rm LR} G^{-1}_{\rm RR} G_{\rm RL} )_{aa} (\omega_{\rm L})^{{1\over2}}_a 
\approx 2 \Sigma_{aa}
\label{altent8}
\eeq
where in arriving at the second equality we have used (\ref{cor5}).
Keeping $\kappa$ arbitrary, we can now use (\ref{ent69}) to write 
$\Sigma_{aa}$ as an integral. This yields
\beq
\P (R| L; a)  \approx
{16 \over \pi^2} {1 \over a} \int_0^{\infty} {d\omega \over \omega} \Delta^2(\omega_{{\rm L}a}, \omega)
\label{altent10}
\eeq
where $\Delta(\omega_{{\rm L}a}, \omega)$ is given in (\ref{ent63}).
While it is difficult to carry out the integration analytically, we can numerically evaluate $\P (R| L; a)$ to bring out the $\kappa$-dependence.
In the graph below we show $\P (R|L; a)$ for
$a = 1$ to $200$, for a few sample values of $(\kappa b/2\pi) = 0, 10, 20, 30, 40, 50$.
The top curve (blue curve) is for $\kappa = 0$ and we move down with
the increasing value of $\kappa$, with the black curve corresponding to
$(\kappa b/2\pi) = 50$.

Since for large values of $a$, or $\omega_{{\rm L}a}\gg \kappa$, the $\kappa$-dependence in $\Delta (\omega_{{\rm L} a},\omega)$ can be neglected, graphs for different $(\kappa b/2\pi)$ should tend towards merging. This is evident from Fig.\,\ref{Fig1}.
Also, for $\omega_{{\rm L}a}\gg \kappa$, we should expect that
$\P (R|L; a)$ should be close to its value at $\kappa = 0$.
This can also be checked numerically. 
For example, for $a = 10^6$, $a\, \P (R|L; a) = 0.0337737, 0.0337721, 0.0337579,
0.0336211$
for $(\kappa b/2\pi) = 0, 10, 100, 1000$, respectively.
Further, $\P (R|L; a) \sim 1/a$ for large values of $a$, in agreement with the
expectation that the trace of $\Sigma$ should diverge logarithmically.

Since for small $\omega$, $\Delta (\omega_{{\rm L} a},\omega) \sim \omega/\kappa $, we expect
$\P (R|L; a)$ to be cut off for small values of $a$. 
This is also clear from 
Fig.\,\ref{Fig1}. This is along the lines of why we expect entanglement to vanish as $\kappa \rightarrow \infty$.

%%%%%%%%%%%%%%%%%%%%%%%%%%%%%%%%%%%%%%%
%%%%%%%%%%%%%%%%%%%%%%%%%%%%%%%%%%%%%%%
\section{Discussion}
%%%%%%%%%%%%%%%%%%%%%%%%%%%%%%%%%%%%%%%
%%%%%%%%%%%%%%%%%%%%%%%%%%%%%%%%%%%%%%%
As mentioned in the Introduction, one can envisage physical situations where entanglement with nontrivial interface conditions for the fields on the entangling surface can be relevant. In particular, the metrical and other geometric data on the interface (both intrinsic and extrinsic)
are expected to be relevant for gravity.
In this paper, we have initiated the analysis of how interface conditions can influence entanglement.
We have considered general Robin conditions
characterized by a parameter $\kappa$.
Our calculational procedure is very direct, integrating out the fields
in part of space in the vacuum pure state density matrix
$\Psi_0 \Psi_0^*$ of the theory.
It is then fairly easy to see how the Reeh-Schlieder theorem, 
which is what encodes the key feature of entanglement, can be 
obtained for $\kappa = 0$. We expect that the same result will hold for
any finite value of the Robin parameter
$\kappa$; this will be analyzed in future.

The entanglement entropy for $\kappa = 0$ reproduces well-known features,
a logarithmic divergence in 1+1 dimensions, equivalent to an area law in higher dimensions.
We have also calculated the correction to the entanglement
entropy for small values of $\kappa$.
The dependence of the entropy on $\kappa$, for large values of $\kappa$,
can be obtained numerically, see Fig.\,\ref{Fig1a}. It is seen to fit well to a logarithmic function,
the specific form has been given in (\ref{ent76d}).
As explained there, this is in accordance with the fact that we have a logarithmic UV divergence.

Since the signature of entanglement is the existence of correlations, even over spacelike separations, there is naturally a relation between the reduced density matrix and the correlation functions or propagators.
The relation between the two, and the possible reconstruction of
 $\rho_{\rm red}$ from correlators was worked out in section 4.

We also introduced a different measure of entanglement which is easier to analyze for general values of $\kappa$. This is the conditional probability 
for detecting a particle in any state  anywhere in one region of space
given that it is introduced in some specific state in the complementary region of space. This conditional probability is also closely related to 
the reduced density matrix.
In Fig.\,\ref{Fig1}, we have shown how this probability changes with
$\kappa$.

As mentioned before, this paper is a first foray into this complex topic.
Further analysis, especially the dependence on more general geometric data, will be taken up in future. But even within the class of Robin conditions, there are further questions of interest.
For example, although 
we have restricted our analysis to positive values of the Robin 
parameter $\kappa$, it is interesting to consider negative values as well.
For negative values of $\kappa$,
there can be bound states corresponding to negative eigenvalues
for the Laplacian, which are localized near the interface \cite{Asorey}.
This is in accordance with the general fact that there can be boundary conditions allowed by the von Neumann theory of self-adjoint 
extensions which allow for negative eigenvalues.
Specifically for our case, if we take
$\omega L = \pm i \lambda$ (so that $\omega^2 <0$), the eigenvalue
condition (\ref{ent8}) becomes
\beq
\lambda\, \sinh \lambda + (\kappa L) \, \sinh (\lambda \gamma) \, \sinh(\lambda (1-\gamma)) = 0,
\hskip .2in \gamma = b/L
\label{disc1}
\eeq
There is no solution for $\lambda$ for $\kappa \geq 0$, but one can have
a nontrivial solution for $\kappa < 0$ for some choices of
$\gamma$.
Being bound states localized near the interface, the corresponding modes
can affect the entanglement and the expression for entropy.
Notice also that the integral expressions for
$\Delta_{\rm L}$, $\Delta_{\rm R}$, $\Delta_{\rm LR}$ in
(\ref{ent60}) involve the denominator
$f(x, \kappa) =  x \bigl[ \coth xb + \coth x (L-b) \bigr] + \kappa$.
For negative values of $\kappa$ one can encounter singularities 
in the integration over $x$, signaling some instability. Clearly this is an interesting question and needs to be investigated further. 

\bigskip
This work was supported in part by the U.S. National Science Foundation Grant No. PHY-2412479 and by the Mina Rees Dissertation Fellowship at the CUNY Graduate Center.

%%%%%%%%%%%%%%%%%%%%%%%%%%%%%%%%%%%%%%%
%%%%%%%%%%%%%%%%%%%%%%%%%%%%%%%%%%%%%%%
%%%%%%%%%%%%%%%%%%%%%%%%%%%%%%%%%%%%%%%
%%%%%%%%%%%%%%%%%%%%%%%%%%%%%%%%%%%%%%%
\section*{Appendix}
\def\theequation{A\arabic{equation}}
\setcounter{equation}{0}
%%%%%%%%%%%%%%%%%%%%%%%%%%%%%%%%%%%%%%%
%%%%%%%%%%%%%%%%%%%%%%%%%%%%%%%%%%%%%%%
%%%%%%%%%%%%%%%%%%%%%%%%%%%%%%%%%%%%%%%
%%%%%%%%%%%%%%%%%%%%%%%%%%%%%%%%%%%%%%%
A related but somewhat different method 
for the calculation
of correlations has been developed in the context of Casimir energy calculations
in \cite{casimir}, incorporating the full generality of boundary conditions
allowed by the von Neumann theorem on self-adjoint extensions.
This has also been used for relating the so-called contact term
in the entanglement for gauge theories in 2+1 dimensions
to the Burghlea-Friedlander-Kappeler formula \cite{AKN-ent}.
It is therefore useful to relate the direct calculation of
correlations presented in section 4.2 to this approach.

We start with the Euclidean version of the action given in (\ref{ent1}).
It is given by
\beq
S_{\rm E} = {1\over 2} \int d^4x\, (\del \phi)^2  + {1\over 2} \int_{x_1=b}
d^3x\,
\vf \, \kappa \, \vf
\label{A1}
\eeq
We have written this in four dimensions to show how the splitting of the field can be done in general, we will return to 1+1 dimensions shortly.
The interface is at $x_1 = b$.

The field $\phi(x)$ defined over the interval $x_1 \in [0,L]$ can be split up as follows:
\beq
\phi(x) = \begin{cases}
\eta_{\rm L}(x) - \int_z n\cdot \del_z G^{\rm L}(x,z) \vf_0 (z)\hskip .2in&x_1\in [0,b]\\
\eta_{\rm R}(x) + \int_z n\cdot \del_z G^{\rm R} (x,z) \vf_0 (z) &x_1\in [b,L]\\
\end{cases}
\label{A2}
\eeq
Here $\eta_{\rm L}$ is the field on the left region obeying Dirichlet conditions, i.e., vanishing, on the interface at $x_1 = b$, as well as
at $x_1 = 0$. Thus it can be expanded
in terms
of the modes $v_a (x)$. Similarly $\eta_{\rm R}$ is defined on the right region
wand vanishes on the interface as well as at $x= L$, expandable in terms
of $w_\alpha$. 
$\vf_0$ is the value of the field $\phi$ on the interface.
In (\ref{A2}), we have continued $\vf_0$ into the left and right regions
such that it obeys the Laplace equation,
\beq
\del^2_x \left[ \int_z  n\cdot \del_z G^{\rm L}(x,z) \vf_0 (z)\right]
= \del^2_x \left[ \int_z n\cdot \del_z G^{\rm R} (x,z) \vf_0 (z)\right] = 0
\label{A3}
\eeq
The continuation has been carried out using Green's theorem
and the Dirichlet Green's functions $G^{\rm L}$, $G^{\rm R}$, obeying
\beq
- \left( \del_1^2 + \nabla^2 \right) \, G(x, x') = \delta^{(4)} (x, x')
\label{A4}
\eeq
where $\nabla^2 = \del_2^2 + \del_3^2 +\del_4^2$. 
Also, in (\ref{A2}) and (\ref{A3}), $n\cdot \del$ denotes the derivative with respect to the 
coordinate normal to the interface, $n^\mu$ being the unit normal vector.

With Dirichlet boundary on
the interface at $x_1 = b$, the Green's function for the left region is
\beqar
G^{\rm L} (x, y) &=& {2\over b} \int {d^3k \over (2\pi)^3} e^{- i \vk\cdot(\vx-\vy)}
\sum_a {\sin (\omega_a x) \sin (\omega_a y) \over k^2 + \omega_a^2}
\nonumber\\
&=& \int {d^3k \over (2\pi)^3} e^{- i \vk\cdot(\vx-\vy)} \, G^{\rm L} (\omega, x, y)
\label{A5}\\
G^{\rm L} (\omega, x, y)&=&
{1\over 2 \omega} \Biggl[ \Theta (x_1 - y_1) \left\{ {e^{\omega (x_1 - y_1)}\over
e^{2 b \omega}- 1} +{e^{-\omega (x_1 - y_1)}\over
1-e^{-2 b \omega}} \right\}\nonumber\\
&& \hskip .25in + \Theta (y_1 - x_1) \left\{ {e^{-\omega (x_1 - y_1)}\over
e^{2 b \omega}- 1} + {e^{\omega (x_1 - y_1)}\over
1- e^{-2 b \omega}} \right\}\nonumber\\
&&\hskip .25in- {e^{\omega (x_1 + y_1)}\over
e^{2 b \omega}- 1} - {e^{-\omega (x_1 + y_1)}\over
1- e^{-2 b \omega}}\Biggr]
\label{A6}
\eeqar
where $\omega_a = (a \pi/b)$, 
$\omega = \sqrt{k_2^2 + k_3^2 + k_4^2}$ and $\Theta (x_1- y_1)$ is the step-function.
 In a similar way, the
Green's function for the right region is given by
\beqar
G^{\rm R} (x, y) &=& {2\over b} \int {d^3k \over (2\pi)^3} e^{- i \vk\cdot(\vx-\vy)}
\sum_\a {\sin (\omega_\a x) \sin (\omega_\a y) \over k^2 + \omega_\a^2}
\nonumber\\
&=& \int {d^3k \over (2\pi)^3} e^{- i \vk\cdot(\vx-\vy)} \, G^{\rm R} (\omega, x, y)
\label{A7}\\
G^{\rm R} (\omega, x, y)&=&
{1\over 2 \omega} \Biggl[ \Theta (x_1 - y_1) \left\{ {e^{\omega (x_1 - y_1)}\over
e^{2 \omega (L-b)}- 1} +{e^{-\omega (x_1 - y_1)}\over
1-e^{-2 \omega (L-b)}} \right\}\nonumber\\
&& \hskip .25in + \Theta (y_1 - x_1) \left\{ {e^{-\omega (x_1 - y_1)}\over
e^{2 \omega}(L-b) - 1} + {e^{\omega (x_1 - y_1)}\over
1- e^{-2 \omega (L-b)}} \right\}\nonumber\\
&&\hskip .25in- {e^{\omega (x_1 + y_1 - 2b)}\over
e^{2 \omega (L-b)}- 1} - {e^{-\omega (x_1 + y_1 - 2b)}\over
1- e^{-2 \omega (L-b)}}\Biggr]
\label{A8}
\eeqar
where $\omega_\a = (\a \pi /(L-b))$.

Using the parametrization in (\ref{A2}), the Euclidean action is seen to be
\beqar
S &=& \int \left[{1\over 2} (\del \eta_{\rm L})^2 + {1\over 2} (\del \eta_{\rm R})^2 
+ {1\over 2} \vf_0 (z) (M_{\rm L} + M_{\rm R} + \kappa )_{z,z'} \, \vf_0 (z')\right]
\label{A9}\\
M_{\rm L} (z, z') &=& - n\cdot \del_{z'} n\cdot \del_{z} G^{\rm L} (z', z)\big\vert_{z_1 = z'_1 = b} \nonumber\\
&=&\int {d^3k\over (2\pi)^3} e^{ -i \vk\cdot (\vz - \vz' )}
 \omega \coth (\omega b )\nonumber\\
M_{\rm R} (z, z') &=& - n\cdot \del_{z'} n\cdot \del_{z} G^{\rm R} (z', z)\big\vert_{z_1 = z'_1 = b} \nonumber\\
&=& \int {d^3k\over (2\pi)^3} e^{ -i \vk\cdot (\vz - \vz' )}\omega \coth (\omega (L-b))
\label{A10}
\eeqar
where $\vz = (z_2, z_3, z_4)$, $\vk = (k_2, k_3, k_4)$, etc.
and $M_{\rm L} (z, z')$ and $M_{\rm R} (z, z')$ are the Dirichlet-to-Neumann kernels for this problem. Using (\ref{A6}) and (\ref{A8}), it is easy to verify that
\beqar
n\cdot \del_{z} G^{\rm L} (x, z )&=& \int {d^3 k \over (2\pi)^3} e^{- i \vk\cdot(\vx - \vz)}\left[- {\sinh \omega x_1 \over \sinh \omega b}\right]\nonumber\\
n\cdot \del_{z} G^{\rm R} (z, y)&=& \int {d^3 k \over (2\pi)^3} e^{- i \vk\cdot(\vx - \vz)} \left[ {\sinh \omega (L- y_1) \over \sinh \omega (L-b)}\right]
\label{A11}
\eeqar

We will now return to 1+1 dimensions, so that only $k_4$, $x_4$, $z_4$
remain in the expressions
(\ref{A4})-(\ref{A11}); the spatial coordinates will be denoted by
$x$, $y$, $z$ for the coordinates
$x_1$, $y_1$, $z_1$, respectively.
From (\ref{A9}), we see that the propagators for
$\eta_{\rm L}$, $\eta_{\rm R}$ are given by the Dirichlet
Green's functions (\ref{A5}), (\ref{A7}), while the interface term 
in (\ref{A9}) determines the
propagator for $\vf_0$.
It is now straightforward to calculate the correlation functions. We find
\beqar
G^{\rm LL} (x, y) &=& \Big\langle \left[ \eta_{\rm L} (x) - \int_z n\cdot \del_z G^{\rm L} (x, z) \vf_0 (z) \right] \left[ \eta_{\rm L} (y) - \int_{z'} n\cdot \del_{z'} G^{\rm L} (y, z') \vf_0 (z') \right]\Big\rangle\nonumber\\
&=&\la \eta_{\rm L} (x) \eta_{\rm L}(y) \ra + \int_{z, z'}
n\cdot \del_z G^{\rm L} (x, z) n\cdot \del_{z'} G^{\rm L} (y, z')\,
\la \vf_0 (z) \vf_0 (z')\ra\nonumber\\
&=&G^{\rm L} (x, y) + \int {d k_4 \over 2\pi} {\sinh(\omega x) \sinh (\omega y) \over (\sinh(\omega b))^2} {e^{-i k_4 (x_4- y_4)}\over \omega \left\{ \coth \omega b + \coth\omega (L-b)\right\} + \kappa } \nonumber\\
&=&G^{\rm L} (x, y) + \int {d k_4 \over 2\pi} {\sinh(\omega x) \sinh (\omega y) \over (\sinh(\omega b))^2} e^{-i k_4 (x_4- y_4)}\, {1\over f(\omega, \kappa )}
\label{A12}
\eeqar
where $\omega$ is now just $\sqrt{k_4^2}$ and $f(\omega, \kappa )$
is as given in (\ref{ent61}). 
Taking the equal-time limit, which is what is relevant for
our purpose, we find
\beq
G^{\rm LL} (x, y) = G^{\rm L} (x, y) + \int_0^\infty {d\omega \over \pi} {\sinh(\omega x) \sinh (\omega y) \over (\sinh(\omega b))^2} {1\over f (\omega , \kappa)}
\label{cor15}
\eeq
Similarly, we get
\beqar
G^{\rm RR} (x, y) &=& G^{\rm R} (x, y)
+\int_0^\infty {d\omega \over \pi} {\sinh(\omega (L-x)) \sinh (\omega (L-y)) \over \left[\sinh(\omega (L-b))\right]^2} \, {1\over f(\omega, \kappa )}
\label{cor16}\\
G^{\rm LR} (x, y) &=& \int_0^\infty {d\omega \over \pi} {\sinh(\omega x) \sinh (\omega (L-y)) \over \left[\sinh (\omega b) \sinh(\omega (L-b))\right]} \, {1\over f(\omega, \kappa )}
\label{cor17}
\eeqar
Using
\beq
\int_0^b  dx \, \sin((\omega_{\rm L})_a  x) \sinh (\omega x)
= (-1)^{a+1} {(\omega_{\rm L})_a \sinh( \omega b) \over \omega^2 + (\omega_{\rm L})_a^2}
\label{cor18}
\eeq
and a similar expression for the integral from $b$ to $L$, we find
\begin{align}
(G_{\rm LL})_{a b} &= {1\over 2 (\omega_{\rm L})_a} \delta_{a b }
+ (-1)^{a+b} {2\over b}  
\int_0^\infty {d x \over \pi}  {(\omega_{\rm L})_a\over (x^2 + (\omega_{\rm L})_a^2)} {(\omega_{\rm L})_b\over
(x^2 + (\omega_{\rm L})_b^2)} \, {1\over f(x, \kappa )}
\nonumber\\
&\equiv{1\over 2 (\omega_{\rm L})_a} \delta_{ac }
+ ({\tilde \Delta}_{\rm L} )_{a c}\label{cor19}\\
(G_{\rm RR})_{\a \b} &= {1\over 2 (\omega_{\rm R})_\a} \delta_{\a \b}
+ (-1)^{\a+ \b} {2 \over L-b}  
\int_0^\infty {d x \over \pi}  {(\omega_{\rm R})_\a \over (x^2 + (\omega_{\rm R})_\a^2)}
{(\omega_{\rm R})_\b \over (x^2 + (\omega_{\rm R})_\b^2)} \, {1\over f(x, \kappa )}
\label{cor20}\\
(G_{\rm LR})_{a\a} &= (-1)^{a+\a} \sqrt{2 \over b} \sqrt{2 \over L-b}  
\int_0^\infty {d x \over \pi}  {(\omega_{\rm L})_a\over (x^2 + (\omega_{\rm L})_a^2)}
{(\omega_{\rm R})_\a \over (x^2 + (\omega_{\rm R})_\a^2)} \, {1\over f(x, \kappa )}
\label{cor21}
\end{align}
These expressions agree with the corresponding formulae in
(\ref{cor6d}), (\ref{cor6e}) and (\ref{cor6f}).

%%%%%%%%%%%%%%%%%%%%%%%%%%%%%%%%%%%%%%%
%%%%%%%%%%%%%%%%%%%%%%%%%%%%%%%%%%%%%%%
%%%%%%%%%%%%%%%%%%%%%%%%%%%%%%%%%%%%%%%
%%%%%%%%%%%%%%%%%%%%%%%%%%%%%%%%%%%%%%%
%%%%%%%%%%%%%%%%%%%%%%%%%%%%%%%%%%%%%%%
%%%%%%%%%%%%%%%%%%%%%%%%%%%%%%%%%%%%%%%
%%%%%%%%%%%%%%%%%%%%%%%%%%%%%%%%%%%%%%%
%%%%%%%%%%%%%%%%%%%%%%%%%%%%%%%%%%%%%%%

%%%%%%%%%%%%%%%%%%%%%%%%%%%%%%%%%%%%%%%%%%%%%%%
%%%%%%%%%%%%%%%%%%%%%%%%%%%%%%%%%%%%%%%%%%%%%%%

%%%%%%%%%%%%%%%%%%%%%%%%%%%%%%%%%%%%%%%%%%%%%%%
%%%%%%%%%%%%%%%%%%%%%%%%%%%%%%%%%%%%%%%%%%%%%%%
%%%%%%%%%%%%%%%%%%%%%%%%%%%%%%%%%%%%%%%%%%%%%%%
%%%%%%%%%%%%%%%%%%%%%%%%%%%%%%%%%%%%%%%%%%%%%%%
%%%%%%%%%%%%%%%%%%%%%%%%%%%%%%%%%%%%%%%%%%%%%%%
%%%%%%%%%%%%%%%%%%%%%%%%%%%%%%%%%%%%%%%%%%%%%%%
%%%%%%%%%%%%%%%%%%%%%%%%%%%%%%%%%%%%%%%%%%%%%%%
%%%%%%%%%%%%%%%%%%%%%%%%%%%%%%%%%%%%%%%%%%%%%%%

\begin{thebibliography}{99}
%%%%%%%%%%%%%%%%%%%%%%%%%%%%%%%%%%%%%%%%%%%%%%%
%%%%%%%%%%%%%%%%%%%%%%%%%%%%%%%%%%%%%%%%%%%%%%%
\bibitem{reviews} 
It is difficult to have a comprehensive list of references for this broad area of work. For some recent reviews, see
T. Nishioka, 
%“Entanglement entropy: holography and renormalization group,” 
Rev. Mod. Phys. {\bf 90} no. 3, (2018) 035007, arXiv:1801.10352 [hep-th];
  H.~Casini and M.~Huerta,
  %``Entanglement entropy in free quantum field theory,''
  J.\ Phys.\ A {\bf 42}, 504007 (2009)
  [arXiv:0905.2562 [hep-th]];
  For a more recent review by the same authors, see
  arXiv:2201.13310;
  P.~Calabrese and J.~Cardy,
  %``Entanglement entropy and conformal field theory,''
  J.\ Phys.\ A {\bf 42}, 504005 (2009), 
  doi:10.1088/1751-8113/42/50/504005
  [arXiv:0905.4013 [cond-mat.stat-mech]].
  
\bibitem{top-phases} This is also a very vast topic, for a review, see
X.G. Wen, \RMP {\bf 89}, 041004 (2017).

\bibitem{q-comp} The literature on this is enormous. To include a few review examples, M.A. Nielsen, I. Chuang, {\it Quantum Computation and Quantum Communication}, Cambridge University Press (2000); I. Bengtsson, K. \.Zyczkowski, {\it Geometry of Quantum States - An Introduction to Quantum Entanglement}, Cambridge University Press (2006);
D. Bru{\ss}, J. Math. Phys {\bf 43}, 4237 (2002);
V. Vedral, Rev. Mod. Phys. {\bf 74}, 197 (2002);
R. Horodecki, P. Horodecki, M. Horodecki, K. Horodecki, Rev. Mod. Phys. {\bf 81}, 865 (2009).

\bibitem{bek-hawk} J.D. Bekenstein, Lett. Nuovo Cim. {\bf 4}, 737 (1972); Phys. Rev. {\bf D7}, 2333 (1973);
S.W. Hawking, Nature {\bf 248}, 30 (1974); Commun. Math. Phys. {\bf 43}, 199 (1975).

\bibitem{jacobson} T. Jacobson, \PRL~{\bf 75}, 1260 (1995); \PRL~{\bf 116}, 201101 (2016);
T. Padmanabhan, Rep. Progr. Phys. {\bf 73}, 6901 (2010);
E. Verlinde, \JHEP 1104:029 (2011).


\bibitem{EE-grav-early} L. Bombelli, R.K. Koul, J. Lee and R.D. Sorkin, Phys. Rev. {\bf D34}, 373 (1986);
G. 't Hooft, Nucl. Phys. {\bf B256}, 727 (1985).

\bibitem{EE-grav-later} M. Srednicki, Phys. Rev. Lett. {\bf 71}, 666 (1993);
V. Frolov and I. Novikov, Phys. Rev. {\bf D48}, 4545 (1993);
D. Kabat and M. Strassler, Phys. Lett. {\bf B329}, 46 (1994);
D.~N.~Kabat,
  %``Black hole entropy and entropy of entanglement,''
  Nucl.\ Phys.\ B {\bf 453}, 281 (1995),
  doi:10.1016/0550-3213(95)00443-V
  [hep-th/9503016];
For a review with references to earlier work: S.N. Solodukhin, Living Rev. Relativ. {\bf 14}, 8 (2011).

\bibitem{EE-grav-string} S. Ryu and T. Takayanagi, Phys. Rev. Lett. {\bf 96}, 181602 (2006); \JHEP~0608:045 (2006).

\bibitem{Raam} See for example, T. Faulkner {\it et al}, arXiv:1312.7856;
 B. Swingle and M. van Raamsdonk, arXiv:1405.2933.

\bibitem{RS} H. Reeh and S. Schlieder, Nuovo Cim. {\bf 22}, 1051 (1961).

\bibitem{vN} H. Araki, Prog. Theoret. Phys. {\bf 32}, 956 (1964); A. Connes and E. Stormer, J. Funct. Analysis, {\bf 28}, 187 (1978);
H.J. Borchers, J. Math. Phys. {\bf 41}, 3604 (2000);
J. Sorce, Rev. Math. Phys., 2430002 (2024).

\bibitem{vN2} R. Clifton and H. Halverson, Stud. Hist. Philos. Mod. Phys. {\bf 32}, 1 (2001); H. Halverson and M. Mueger, arXiv:math-ph/0602036;
S. Hollands and K. Sanders, arXiv:1702.04924[quant-ph];
C.J. Fewster and K. Rejzner, arXiv:1904.04051.

\bibitem{witten} For a review of entanglement in the context of operator algebras, see E. Witten, \RMP~{\bf 90}, 045003 (2018) (arXiv: 1803.04993[hep-th]).

\bibitem{replica} L. Susskind, arXiv:hep-th/9309145; C. Callan and F. Wilczek, Phys. Lett. {\bf B333}, 55 (1994); C. Holzhey, F. Larsen and F. Wilczek, Nucl. Phys. {\bf B424}, 443 (1994), see also
\cite{EE-grav-later}.

\bibitem{Asorey}
J. von Neumann, Math. Ann. {\bf 102} 49 (1929);
M. Asorey, A.Ibort and G. Marmo, 
%``Global theory of quantum boundary conditions and topology change", 
Int. J. Mod. Phys.  {\bf A20}  1001 (2005);
M. Asorey, D. Garcia-Alvarez, J. M. Munoz-Castaneda,
%``Casimir Effect and Global Theory of Boundary Conditions", 
J. Phys. {\bf A39} 6127 (2006);
M. Asorey, J. M. Munoz-Castaneda, 
%``Vacuum Boundary Effects",
J. Phys. {\bf A41} 304004 (2008).

\bibitem{casimir} D. Kabat, D. Karabali and V.P. Nair,
Phys. Rev. {\bf D81}, 125013 (2010); Phys. Rev. {\bf D84}, 129901 (2011) (Erratum); Phys. Rev. {\bf D82}, 025014 (2010);
D. Karabali and V.P. Nair, Phys. Rev. {\bf D87}, 105021 (2013).

\bibitem{AKN-ent} A. Agarwal, D. Karabali and V.P. Nair, Phys. Rev. {\bf D96}, 125008 (2017).










%%%%%%%%%%%%%%%%%%%%%%%%%%%%%%%%%%%%%%%%%%%%%%%
 %%%%%%%%%%%%%%%%%%%%%%%%%%%%%%%%%%%%%%%%%%%%%%%
%%%%%%%%%%%%%%%%%%%%%%%%%%%%%%%%%%%%%%%%%%%%%%%
\end{thebibliography}
\end{document}